\documentclass[10pt,a4paper,onecolumn,final]{article}

\usepackage{amsmath}

\usepackage{amssymb}

\usepackage[final]{graphicx}

\begin{document}

\title{ANOMALOUS FLUXON PROPERTIES OF LAYERED SUPERCONDUCTORS}

\author{V.M.Krasnov \\
\it Department of Microelectronics and Nanoscience, Chalmers \\
\it University of Technology, S-41296 G\"oteborg, Sweden}

%\date{\today }

\maketitle

%\vspace{120pt}

\section{INTRODUCTION}

\ \ \ \ \ Highly anisotropic layered superconductors can be
considered as stacks of Josephson junctions (SJJ's)
\cite{LD,Muller}. The key feature of SJJ's is mutual coupling of
junctions, which can be achieved via magnetic \cite{Mineev,SBP},
charging \cite{charging_Tachiki,charging_Preis} or non-
equilibrium \cite{Ryndyk} coupling mechanisms. Properties of SJJ's
can be qualitatively different from that of single Josephson
junctions (JJ's) or anisotropic type-II superconductors. This is
reflected in a structure of magnetic vortices parallel and
perpendicular to superconducting (S) layers. A perpendicular
vortex represent an array of 2D pancake vortices in S-layers
\cite{Clem1}. Those are Abricosov type vortices with normal cores
and with supercurrents flowing in S-layers at a characteristic
length $\lambda _{ab}$. Properties of such vortices are well
studied, see e.g. Ref. \cite{Blatter}, and will not be considered
here. For parallel or slightly tilted magnetic field, vortex
system commensurates with the layered structure \cite{Nelson} due
to intrinsic pinning \cite{IntPin}, which tends to locate vortices
between S-layers. Such vortices (fluxons) are of Josephson type.
They have circulating currents both along and across layers and do
not have normal cores.

Fluxon related properties of SJJ's were extensively studied both
analytically and numerically [3,4,12-32], especially in connection
with High-$T_c$ superconductors (HTSC). Due to a high anisotropy,
(weak Josephson coupling between layers), even small in-plane
magnetic field causes penetration of fluxons. For Tl and Bi-based
HTSC, the penetration field is $\sim$ few Oe [33-35]. Since fluxon
energy is low, they can be easily created by thermal excitations
\cite{Blatter,Nelson} or trapped during cooling down. Moreover,
because fluxons can not move freely across layers,
fluxon-antifluxon pairs in different junctions are stable and can
exist even at zero magnetic field \cite{Compar}. Clearly, a
knowledge of fluxon structure is necessary for understanding
transport and magnetic properties of layered superconductors. From
application point of view, it is particularly interesting to
consider SJJ's with finite number layers, which is the case for
HTSC mesas [30,36-43] and low-$T_c$ SJJ's [30,44-54].

Here I will review anomalous single and multi-fluxon properties of
SJJ's. Throughout the paper I will compare analytical, numerical
and experimental results both for intrinsic HTSC and artificial
low-$T_c$ SJJ's.

\section{GENERAL RELATIONS}

\ \ \ \ \  The most common type of coupling in SJJ's is magnetic
coupling. It appears when S-layers are thinner than London
penetration depth, so that magnetic induction is not screened in
one junction. SJJ's are coupled via shared magnetic induction and
currents flowing along common electrodes. Properties of
magnetically coupled SJJ's are described by a coupled sine-Gordon
equation (CSGE), which was phenomenologically introduced in Ref.
\cite{Mineev} and later rigorously derived in Ref. \cite{SBP}

First, I will briefly describe the formalism of CSGE, following
notations of Ref. \cite{Modes}. Let's consider a stack of $N$
junctions with the following parameters: $J_{ci}$ -the critical
current density, $C_i$ and $R_i$- the capacitance and the
quasiparticle resistance per unit area, respectively, $t_i$- the
thickness of tunnel barrier, $d_i$ and $\lambda _{Si}$ - the
thickness and London penetration depth of $S-$layers,
respectively, and $L$- the in-plane lengths of the stack. Elements
of the stack are numerated from the bottom to the top, so that JJ
$i$ consists of S-layers $i$, $i+1$ and a tunnel barrier $i$. The
CSGE can be written in a compact matrix form \cite{SBP}:

\begin{equation}
{\bf \varphi }^{\prime \prime }={\bf A\cdot J}_z-{\bf J}_b
\label{Eq.1}
\end{equation}

\noindent Here ${\bf \varphi }$ is a column of gauge invariant
phase differences $\varphi _i$, 'primes' denote in-plane spatial
derivatives, ${\bf A}$ is a symmetric tridiagonal $N\times N$
matrix with nonzero elements: $A_{i,i-1}=-S_i/\Lambda _l$;
$A_{i,i}=\Lambda _i/\Lambda _l$; $A_{i,i+1}=-S_{i+1}/\Lambda _l$,
where $\Lambda _i=t_i+\lambda _{Si}coth\left( \frac{d_i}{\lambda
_{Si}}\right) +\lambda _{Si+1}coth\left( \frac{d_{i+1}}{\lambda
_{Si+1}}\right) $, $S_i=\lambda _{Si}cosech\left(
\frac{d_i}{\lambda _{Si}}\right) $. Current density across layers,
${\bf J}_z$, consists of supercurrent, displacement current and
normal current components:

\begin{equation}
J_{zi}=j_{ci} sin\left( \varphi _i\right) + \tilde C_i%
\ddot \varphi _i+\alpha _i\dot \varphi _i.  \label{Eq.2}
\end{equation}

\noindent Here $j_{ci}=\frac{J_{ci}}{J_{cl}}$, $\tilde
C_i=\frac{C_i}{C_l}$, 'dots' denote time derivatives,
$\alpha_i=\sqrt{\frac{\Phi _0}{2\pi cJ_{cl}C_lR_i^2}}$ are damping
parameters, $\Phi _0$ is the flux quantum and $c$ is the velocity
of light in vacuum. ${\bf J}_b$ is a bias term \cite{Modes}. Space
and time are normalized to Josephson penetration depth, $\lambda _{jl}=%
\sqrt{\frac{\Phi _0c}{8\pi ^2J_{cl}\Lambda _l}}$, and the inverse
plasma frequency $1/\omega _{pl}=\sqrt{\frac{\Phi _0C_l}{%
2\pi cJ_{cl}}}$ of JJ $l$, respectively.

Magnetic induction in a stack is given by:

\begin{equation}
{\bf B}=\frac{H_0}2{\bf A}^{-1}{\bf \varphi }^{\prime }.
\label{Eq.5}
\end{equation}

\noindent Here $H_0=\Phi_0 /(\pi\lambda_{jl}\Lambda_l)$.

\begin{table}[b]
\noindent
\begin{minipage}{0.99\textwidth}
\begin{center}
\caption{Parameters of LD model in terms of CSGE and estimations
for Bi2212 HTSC}
\vspace{6pt}
\begin{tabular}{|cccc|}
\multicolumn{1}{|c|}{LD} & \multicolumn{1}{c|}{$\lambda _{ab}$} &
\multicolumn{1}{c|}{$\lambda _c=\gamma \lambda _{ab}$} &
\multicolumn{1}{c|}{$\lambda _J=\gamma s $} \\ \hline
\multicolumn{1}{|c|}{CSGE} & \multicolumn{1}{c|}{$\lambda
_s\sqrt{\frac sd}$} & \multicolumn{1}{c|}{$\sqrt{\frac{\Phi
_0c}{8\pi ^2J_cs}}$} & \multicolumn{1}{c|}{$\sqrt{\frac{\Phi
_0cd}{16\pi ^2J_c\lambda _s^2}}=\frac{\lambda
_J(\text{LD})}{\sqrt{2}}$} \\ \hline \multicolumn{1}{|c|}{Bi2212}
& \multicolumn{1}{c|}{0.15-0.2 $\mu
m$} & \multicolumn{1}{c|}{130-184 $\mu m$} & \multicolumn{1}{c|}{$\lambda _J($%
CSGE$)=$0.6-1.1 $\mu m$} \\
\end{tabular}
\end{center}
\end{minipage}
\end{table}

For $d/\lambda _s\ll 1$, equations analogous to Eq.(1) were
derived from Lawrence-Doniach (LD) \cite{LD} model
\cite{Clem2,Bul1}. The LD-CSGE conversion is shown in Table 1,
along with estimations for Bi2212 HTSC ($d$=3 \AA , $t$ =12 \AA ,
$J_c$=500-1000 A/cm$^2$, $\lambda _s$=750-1000 \AA).

In static case, CSGE reduces to

\begin{equation}
{\bf \varphi }^{\prime \prime }={\bf A\cdot}j_c sin({\bf\varphi}).
\label{Eq.7}
\end{equation}

For non-dissipative fluxon motion with a velocity $u$, the phase
distribution remains unchanged in a coordinate frame moving along
with the fluxon, $\xi = x-ut$, and CSGE can be simplified
\cite{FluxoN}

\begin{equation}
{\bf D \cdot \varphi }^{\prime \prime }={\bf A\cdot}j_c
sin({\bf\varphi}), \label{Eq.8}
\end{equation}

\noindent where ${\bf D}={\bf E}-u^2 \tilde C {\bf A}$, and ${\bf
E}$ is the unitary matrix.

{\bf The first integral:} CSGE has a first integral \cite{Modes}

\begin{equation}
\frac{1}{2}\varphi^{\prime *}{\bf G}\varphi^{\prime
}+\sum{j_{ci}[cos(\varphi_i)-1]} = {\bf C}, \label{Eq.FI}
\end{equation}

\noindent where ${\bf C}$ is a constant of the first integral,
$\varphi^{\prime *}$ is a string of $\varphi^{\prime}_i$ and a
matrix ${\bf G}={\bf A}^{-1}$ in static case and ${\bf D A}^{-1}$
for soliton-like fluxon motion.

{\bf The free energy} of the stack is a sum of magnetic, kinetic
and Josephson energies. Using the first integral, Eq.(6), a simple
expression for the free energy density is obtained in static case
\cite{Modes}:

\begin{equation}
{\bf F}={\bf C}+2\sum{j_{ci}[1-cos(\varphi_i)]}. \label{Eq.FE}
\end{equation}

\noindent Here $\bf F$ is measured in units of $\Phi_0 J_{cl}
\lambda_J/2\pi c$. From Eq.(7) it follows that the energy of any
isolated solution (${\bf C}=0$) is twice the Josephson energy,
just like in single JJ.

\section{ A SINGLE FLUXON}

\ \ \ \ \ Despite considerable theoretical efforts, there is still
no exact fluxon solution for SJJ's. However, several approximate
solutions were proposed. For infinite layered superconductor
($N=\infty$) the first approximate solution was obtained within LD
model \cite{LD}. It was suggested that far from the fluxon center,
current stream lines are elliptical with different London
penetration lengths along and across the layers
\cite{Bulaevskii73}. The elliptic solution was extended by Clem
and Coffey \cite{Clem2,ClemCoffey91} to explicitly take into
account discreteness of layers and consider fluxon "core" region.
Extension for superconductors with different layers was made in
Ref. \cite{KrGol}. The first approximate solution for finite SJJ's
was derived for the case of two weakly coupled SJJ's
\cite{Malomed}. However, since coupling strength was used as a
perturbation parameter, that solution does not apply for strongly
coupled SJJ's, which is the most interesting case. A
"multi-component" solution, valid for arbitrary double SJJ's was
obtained in Ref. \cite{Modes}, using phase differences in
fluxon-free junctions as a perturbation \cite{SolPert}. Such
solution was shown to be in a good agreement with numerical
simulations both for static and dynamic cases \cite{FluxonD}.
Recently, the "multi-component" solution was extended for SJJ's
with arbitrary number of junctions \cite{Vortex99,FluxoN}.

\begin{figure}
\begin{center}
\includegraphics[width=9.5cm]{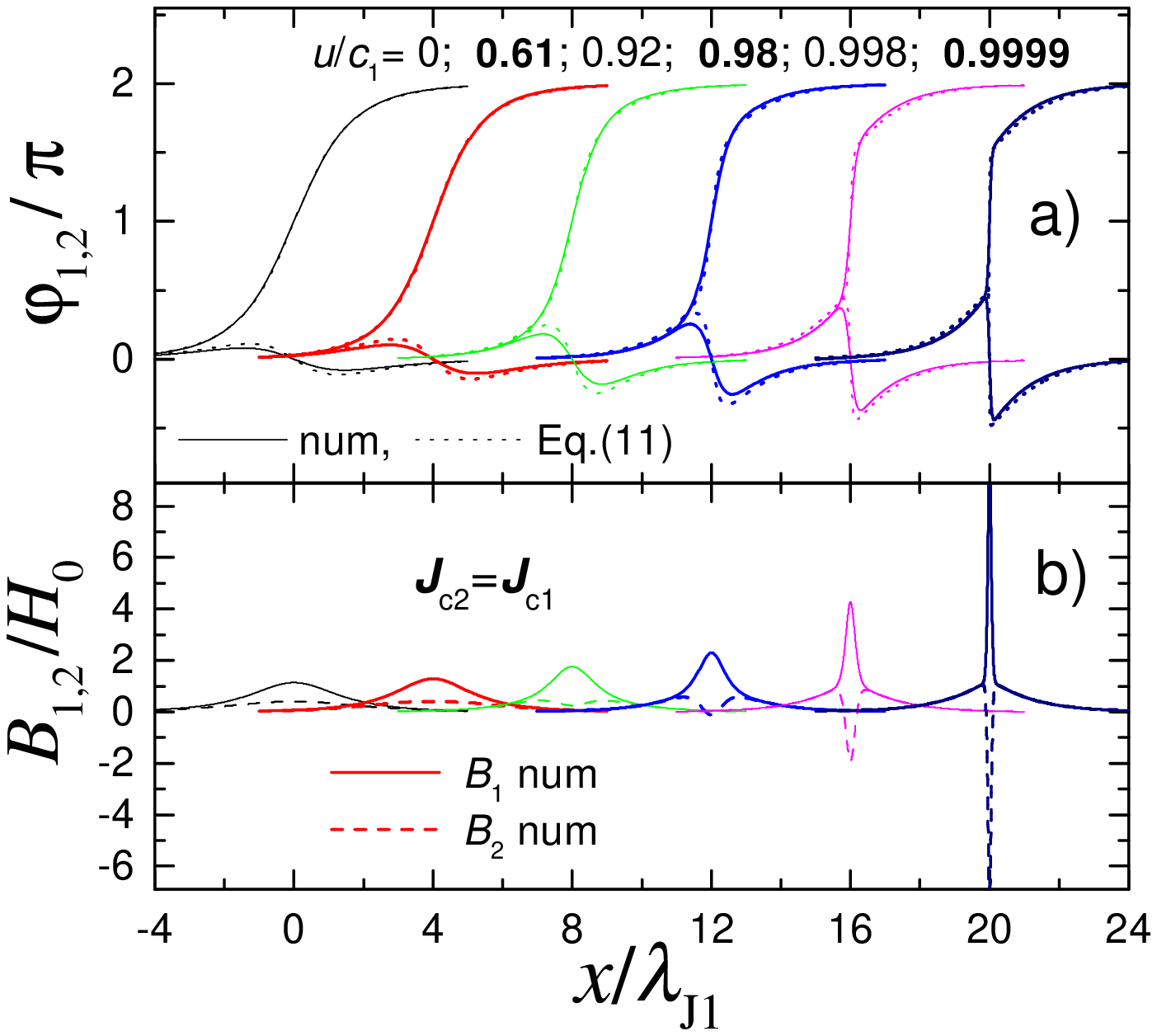}
\caption{ Fluxon shapes in double SJJ's for different fluxon
velocities: a) Solid and dotted lines represent numerical and
analytic solutions, respectively, for $\varphi _{1,2}$. b)
Magnetic inductions $B_{1,2}$ obtained numerically. The existence
of contracted and uncontracted components and the sign inversion
of $B_2(0)$ at $u\rightarrow c_1$ is clearly seen. From Ref.
\cite{SolPert}.} \label{Fig3_SolPer}
\end{center}
\end{figure}

Apart from the multi-component, two other solutions were predicted
in Ref. \cite{Modes} for double SJJ's: (i) the single component
solution, $sin(\varphi_1) \simeq \kappa sin(\varphi_2)$; and (ii)
a combination of the single component solution with a traveling
plasma wave. Numerical modeling showed \cite{FluxonD} that both
solutions can be realized at high fluxon velocities. The latter
solution was independently discovered in Ref. \cite{HechtfPRL} and
was interpreted as "Cherenkov radiation" from a rapidly moving
fluxon.

\subsection{Approximate fluxon solution}

\ \ \ \ \ Let's consider arbitrary stack of $N$ junctions with a
single fluxon in junction $i_0$. The problem with solving CSGE,
Eq.(4), is coupling of nonlinear $sin(\varphi _i)$ terms in the
right-hand side. Partial linearization with respect to $\varphi_{i
\ne i_0}$ was proposed in Refs.\cite{SolPert,FluxoN} for
decoupling of CSGE. First Eq.(4) is diagonalized to

\begin{eqnarray}
\lambda _m^2F_m^{\prime \prime }=sin(F_m)+Er_m, \label{Eq.10} \\
F_m=\varphi _{i_0}+\sum \kappa _{m,i}\varphi _{i\neq i_0},
\label{Eq.11}
\end{eqnarray}

\begin{figure}
\begin{center}
\includegraphics[width=13cm]{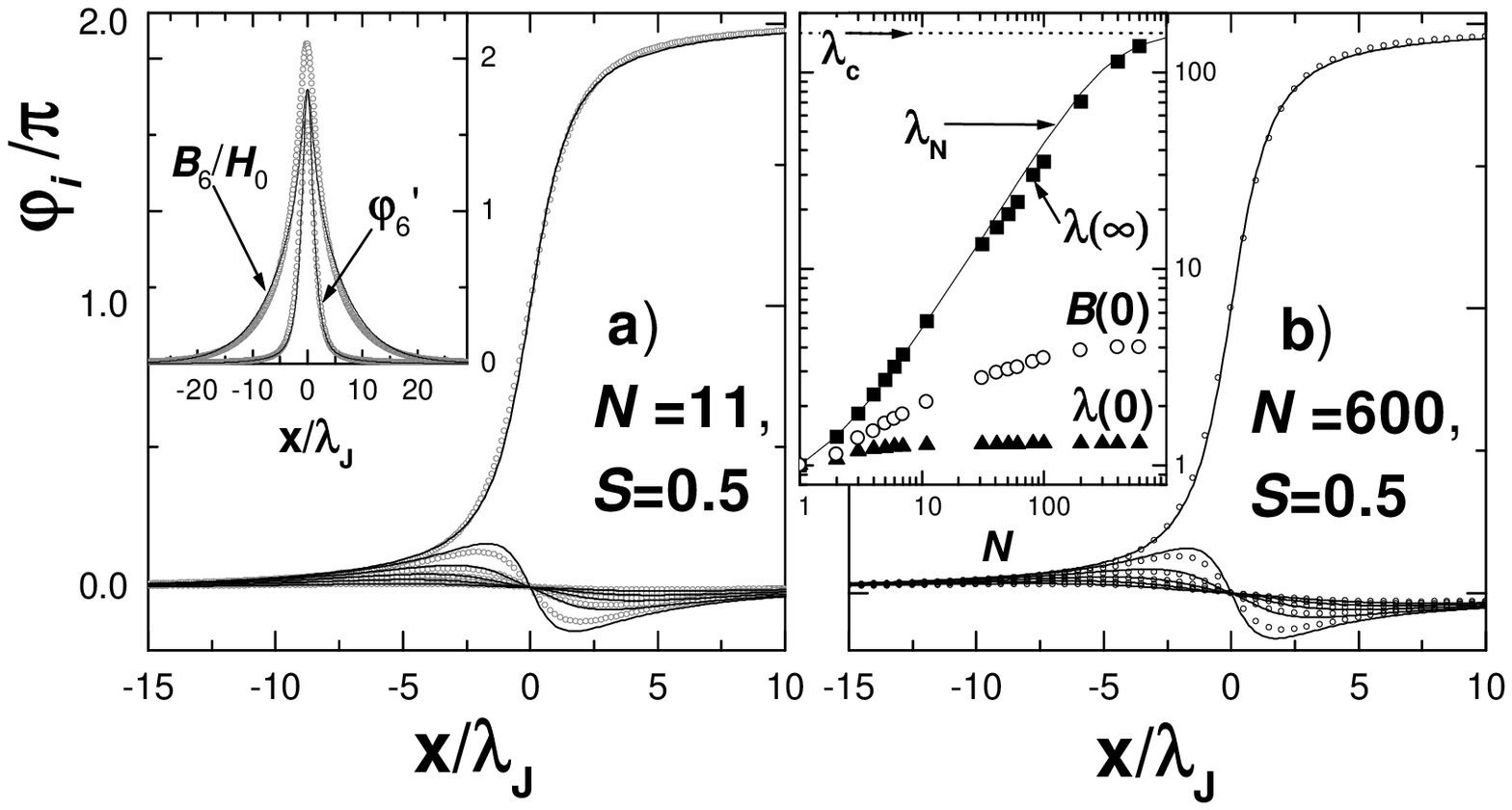}
\vspace*{6pt} \caption{ Phase distribution for a fluxon in a)
$N$=11 and b) 600 SJJ's. Circles and lines represent numerical and
analytic solutions, respectively. Insets show a) variation of
magnetic induction and the derivative of phase in the central JJ;
b) variation of characteristic parameters of the fluxon vs. the
number of SJJ's. Data from Ref. \cite{FluxoN}} \label{Fig.1_2N}
\end{center}
\end{figure}

\noindent where $Er_m=sin(\varphi _{i_0})+\sum \kappa
_{m,i}sin(\varphi _{i\neq i_0}) - sin(\varphi _{i_0}+\sum \kappa
_{m,i}\varphi _{i\neq i_0}) $. Characteristic lengths, $\lambda
_m^{-2}$ and coefficients $\kappa _{m,i}$ are given by eigenvalues
and eigenvectors of ${\bf A}$, respectively. The diagonalization
procedure minimizes error functions, $Er_m$ far from the center.
In this case $Er_m \sim \varphi_i^3$, while any other linear
combination would yield $\sim \varphi_i$. Therefore, $Er_m$ have a
form of small ripple and vanish both inside and far from the
fluxon center. In the first approximation $Er_m$ terms in Eq.(8)
can be neglected, leading to a set of decoupled sine-Gordon
equations for $F_m$ with well known solutions:

\begin{equation}
F_m =4arctan\left( e^{x/\lambda _m}\right) .
\label{Eq.9}
\end{equation}

Finally, inverting Eq.(9) we obtain the approximate fluxon
solution \cite{SolPert,FluxoN} :

\begin{equation}
{\bf \varphi }={\bf K}^{-1}F_m(\lambda _m),  \label{Eq.12}
\end{equation}

\noindent where ${\bf K}$ is the $N\times N$ matrix with elements
equal to $\kappa _{m,i}$. For identical SJJ's eigenvalues/vectors
are given by \cite{Klein1,SakPed},

\begin{eqnarray}
\lambda _m &=&\lambda_J\left[ 1+2Scos\left( \frac{\pi
m}{N+1}\right) \right] ^{-1/2},\ m=1,2, \ldots N,
%\eqnum{13 a}
\label{Eq.13a} \\ \kappa_{m,i} &=&\left( -1\right)
^{i-i_0}\frac{sin\left[ \pi mi/\left( N+1\right) \right]
}{sin\left[ \pi mi_0/\left( N+1\right) \right]. }
%\eqnum{13 b}
\label{Eq.13b}
\end{eqnarray}

\noindent Here $S=S_i/\Lambda _i$ is the coupling parameter,
$S=0\div 0.5$. Normalization of eigenvectors in Eq.(13) follows
from Eq.(9) and ensures that the total phase shift is $2\pi$ for
$i=i_0$ and zero for the rest of the junctions.

For SJJ's with odd $N$ and a fluxon in the middle junction,
$i_0=N/2$, the number of components reduces to $n=(N+1)/2$, due to
a symmetry relation, $\varphi _{i_0-j}=\varphi _{i_0+j}$, and the
solution becomes particularly simple:

\begin{eqnarray}
\varphi _{i_0} &=&\frac {2}{N+1}\sum F_m\text{ }\left( m=1,3,
\ldots N\right),
%\eqnum{14 a}
\label{Eq.14a} \\ B_{i_0} &=
&\frac{H_0}{N+1}\sum \lambda _m^2F_m^{\prime }.
%\eqnum{14 b}
\label{Eq.14b}
\end{eqnarray}

\begin{figure}
\begin{center}
\includegraphics[width=15cm]{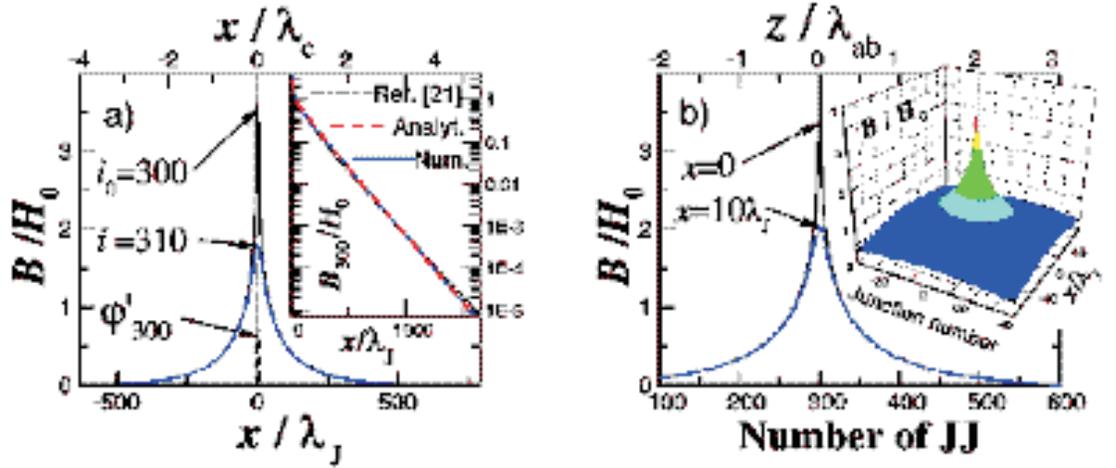}
%\vspace*{6pt}
\caption{Numerically simulated distributions of magnetic induction
for the fluxon in $N$=600 SJJ's: a) $B(z=const)$ along the
$ab$-plane. Dashed line shows the derivative of phase in the
central junction. Inset represents comparison between numerical,
multi-component and elliptic\cite{Clem2} solutions in the central
junction. b) $B(x=const)$ along the $c$-axis. Inset represents a
3D plot of $B(x,z)$ in the fluxon core region. From Ref.
\cite{FluxoN}} \label{fig3N}
\end{center}
\end{figure}

Here are examples of fluxon solutions for small $N$:

\[
N=2:\left\{
\begin{array}{c}
\varphi _1=\frac{F_1+F_2}2,\lambda _1=\lambda _J\left( 1+S\right)
^{-1/2} \\ \varphi _2=\frac{-F_1+F_2}2,\lambda _2=\lambda _J\left(
1-S\right) ^{-1/2}
\end{array}
\right.
\]

\[
N=3:\left\{
\begin{array}{c}
\varphi _2=\frac{F_1+F_3}2,\lambda _1=\lambda _J\left(
1+\sqrt{2}S\right) ^{-1/2} \\
\varphi _{1,3}=\frac{-F_1+F_3}{2\sqrt{2}},\lambda _3=\lambda _J\left( 1-%
\sqrt{2}S\right) ^{-1/2}
\end{array}
\right.
\]

\[
N=5:\left\{
\begin{array}{c}
\varphi _3=\frac{F_1+F_3+F_5}3,\lambda _1=\lambda _J\left( 1+\sqrt{3}%
S\right) ^{-1/2} \\ \varphi
_{2,4}=\frac{-F_1+F_5}{2\sqrt{3}},\lambda _3=\lambda _J \\
\varphi _{1,5}=\frac{F_1-2F_3+F_5}6,\lambda _5=\lambda _J\left( 1-\sqrt{3}%
S\right) ^{-1/2}
\end{array}
\right.
\]

A characteristic feature of the multi-component solution is that
the fluxon is characterized by $N$ lengths, $\lambda _m$,
different from the two lengths of the in elliptic solution
\cite{Bulaevskii73,Clem2,ClemCoffey91}. The approximate fluxon
solution for the dynamic CSGE, Eq.(5), is given by the same
expression, Eq.(11), but with renormalized characteristic lengths,

\begin{equation}
\tilde\lambda_m^2 = \lambda_m^2(1-(\frac{u}{c_m})^2),
%\eqnum{15}
\label{Eq.15}
\end{equation}

\noindent with characteristic velocities \cite{Klein1,SakPed}

\begin{equation}
c_m = c_0\left[ 1+2Scos\left( \frac{\pi m}{N+1}\right)\right]
^{-1/2},\ m=1,2 \ldots N.
%\eqnum{16}
\label{Eq.16}
\end{equation}

\noindent Here $c_0=c\sqrt{t/\varepsilon_r \Lambda}$ is Swihart
velocity for a single junction.

In Fig. 1 spatial distributions of a) phase and b) magnetic
induction are shown for a fluxon in two strongly coupled
($S=0.495$) identical SJJ's at different velocities. Parameters of
the stack are: $d_{1-3}=t_{1,2}=0.01\lambda _{J}$, $\lambda
_{S1-3}=0.1\lambda _{J}$. A quantitative agreement between
''exact'' and approximate solutions is seen. For identical double
SJJ's exactly one half of the fluxon belongs to each of the
components, $F_{1,2}$. Indeed, from Fig. 1 a) it is seen that for
$u\simeq c_1$ there is a Lorentz contracted core at $x=0$ with
one-$\pi $ step in $\varphi _1$. On both sides of the core, there
are two $\pi /2$ tails, decaying at distances $\sim
\widetilde{\lambda }_2\gg \widetilde{ \lambda }_1$. Fig. 1 b)
represent numerically simulated profiles of $B_{1,2}(x)$. It is
seen that the fluxon shape becomes unusual in dynamic state: a dip
in $B_2$ develops in the center of the fluxon with increasing
velocity and at $u\rightarrow c_1$ $B_2(0)$ changes sign. At
$u=c_1$, $B_2(0) = -B_1(0)$, in agreement with the multi-component
solution.

\begin{figure}
\begin{center}
\includegraphics[width=6.5cm]{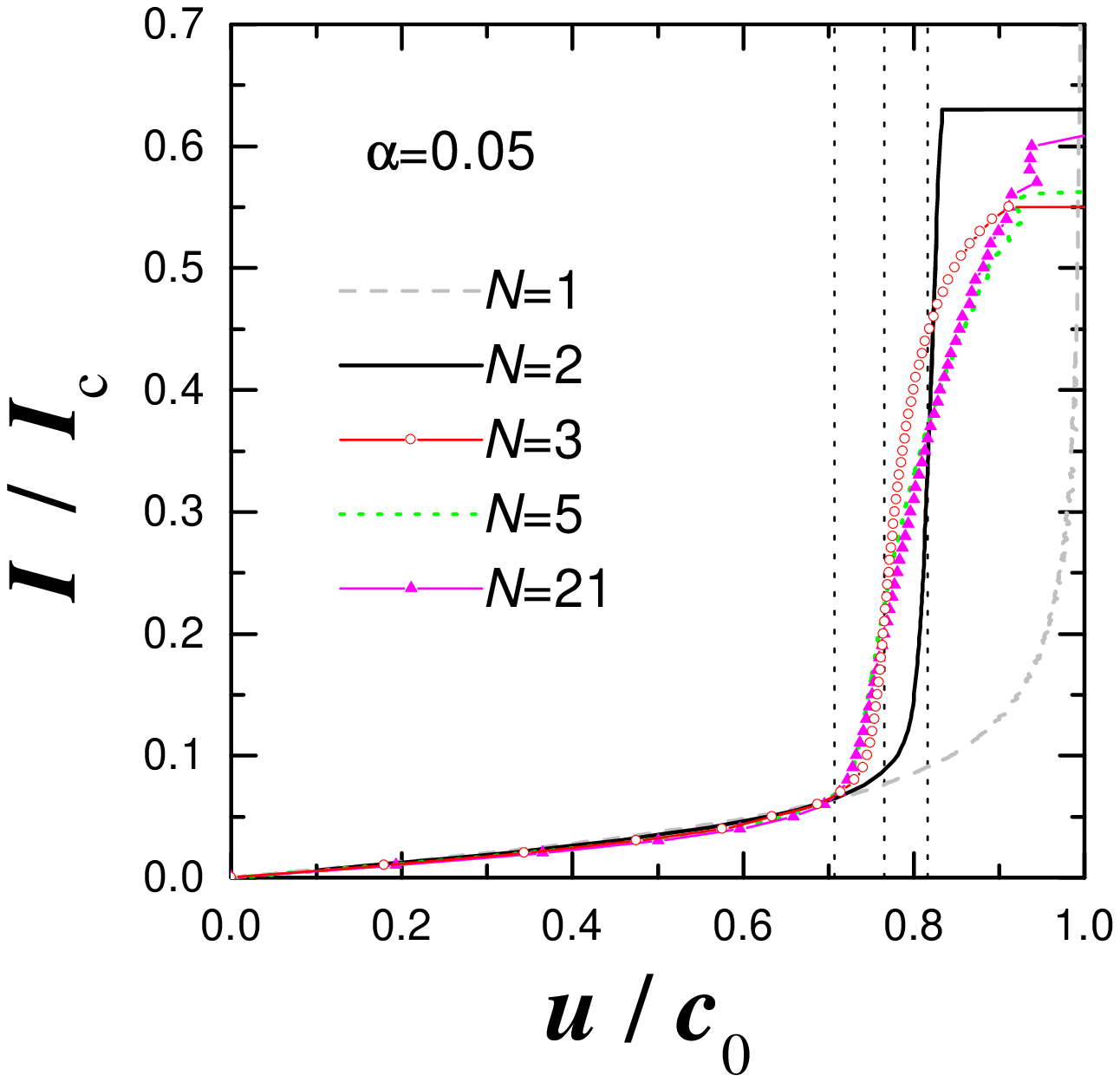}
\caption{Numerically simulated current-velocity characteristics
for a single JJ $N=1$ (dashed line), and for SJJ's with different
$N$: $N=2$ (solid line), $N=3$ (open circles), $N=5$ (dotted line)
and $N=21$ (filled triangles). From Ref.\cite{FluxoN}}
\label{fig7}
\end{center}
\end{figure}

Numerically simulated fluxon profiles for $N=$11 and 600 SJJ's are
shown in Figs. 2 and 3. Parameters are typical for Bi2212: $d$=3
\AA , $t$ =12 \AA , $\lambda_J=1\mu m$ ($J_c \simeq 675 A/cm^2$),
$\lambda _s$=750 \AA ($\lambda_{ab}\simeq 0.17 \mu m$). It was
shown that the multi-component solution correctly describes the
fluxon shape in static case for arbitrary SJJ
\cite{Modes,SolPert,FluxonD,FluxoN}.

\subsection{Decoupling of phase and field}

\ \ \ \ \ Inset in Fig. 2 a) shows spatial variation of $B$ and
$\varphi ^{\prime }$ in the central JJ $i_0$=6 of $N=11$ SJJ's. It
is seen that length scales for variation of $\varphi $ and $B$ are
different: $\varphi $ varies at distances $\simeq \lambda _J$,
while the characteristic length for variation of $B$ at large
distances, $\lambda(\infty)$, is $\lambda _{11} = 5.4\lambda _J$,
Eq.(12). The discrepancy between $\varphi _{i_0}^{\prime }$ and
$B_{i_0}$ becomes dramatic for large $N$, see Fig. 3 a). This is
in sharp contrast with behavior of Josephson vortices in single
JJ's or Abricosov vortices in type-II superconductors, in which
spatial variations of both phase/current and magnetic induction
are given by the same length scale. Existence of one and the same
length scale for $I$ and $B$ is viewed as a natural consequence of
Ampere's law.

So, how could phase/current decouple from magnetic induction in
SJJ's? This can be understood from the multi-component fluxon
solution: (i) Magnetic induction in SJJ's is non-local, $B_i$
depends on $\varphi_i$ in all junctions, see Eq.(3), which is an
essence of magnetic coupling. (ii) If the fluxon has a
multi-component structure, then distribution of phase is
determined by the shortest, while magnetic induction - by the
longest characteristic length. Indeed, according to Eq. (14), the
effective Josephson penetration depth in the center of the fluxon
is

\begin{equation}
\lambda(0)=\frac 2{\varphi _{i_0}^{\prime }(0)}\simeq \frac n{\sum
\lambda _m^{-1}},
%\eqnum{17}
\label{Eq.17}
\end{equation}

\noindent which is obviously dominated by the shortest length
$\lambda _1 \sim \lambda _J$. On the other hand, from Eq. (15),
variation of $B$ is dominated by the longest length, $\lambda _N$,
which according to Eq. (12) increases rapidly with $N$ for
strongly coupled SJJ's, $S \rightarrow 0.5$.

\subsection{N-dependence and assymptotics}

\ \ \ \ \ Fluxon shape strongly depends on the number of junctions
in a stack. Inset in Fig. 2 b) summarizes variation of the fluxon
shape for different $N$. Open circles represent magnetic induction
at the center of the fluxon $B(0)$. $B(0)$ increases with $N$ and
saturates at $B(0)\simeq 4H_0$ for $N>\lambda _{ab}/s$. Triangles
represent numerically obtained $\lambda(0)$, which characterize
variation of phase/current in the fluxon core. In agreement with
Eq. (18), $\lambda(0)$ increases only slightly with $N$ and
saturates at $\simeq 1.3\lambda _J$ for $N>10$. On the other hand,
the effective magnetic length far from the core,
$\lambda(\infty)$, increases dramatically with $N$.
$\lambda(\infty)$ is determined by the largest characteristic
length $\lambda _N$, as shown by the solid line. For large
$N>\lambda _{ab}/s$, $\lambda _N \rightarrow \lambda _c$, see
dotted line in inset of Fig.2 b), and the multi-component solution
approaches the elliptic solutions \cite{Bulaevskii73,Clem2}. This
is demonstrated in inset of Fig. 3 a). Top axes in Figs. 3 a) and
b) show that $B(x,z)$ varies at length scales $\sim \lambda _c$
and $\lambda _{ab}$ along and across layers, respectively. It is
also seen that the most spectacular feature of the fluxon is a
sharp peak $B(0,0)$, representing the ''Josephson core'' of the
fluxon $\sim \lambda _J\times s$ in $ab$-plane and $c$-axis,
respectively, as shown in inset of Fig. 3 b).

\begin{figure}
\begin{center}
\includegraphics[width=6.5cm]{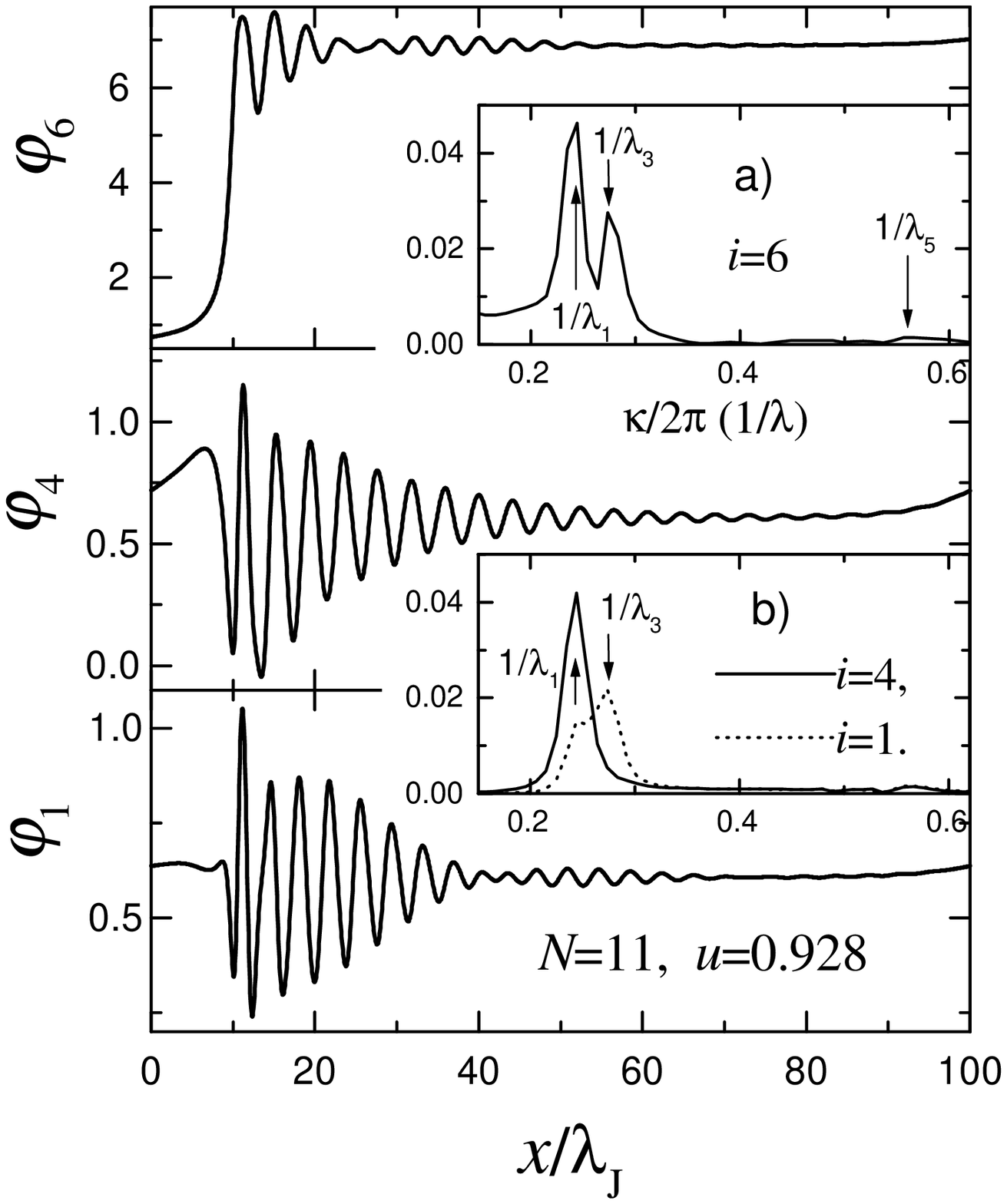}
\caption{Phase distributions for the fluxon moving with velocity
$u\simeq 0.928 c_0$ in $N=11$ SJJ's are shown for the central
junction, $i=6$, (top panel); $i=4$ (middle panel); and $i=1$
(bottom panel). Spectra of Cherenkov radiation are shown in insets
a) for $i=6$ and b) for $i=4$ (solid line) and $i=1$ (dotted
line). From Ref. \cite{FluxoN}. } \label{fig8N}
\end{center}
\end{figure}

Since the fluxon has several characteristic lengths, a question
may arise, what is the effective Josephson penetration depth,
$\lambda _{J(eff)}$. This is important to know, because behavior
of SJJ's changes drastically when $L
> \lambda _{J(eff)}$. Roughly speaking, fluxons can exist only in
"long" JJ's, $L\gg\lambda _{J(eff)}$, causing a large difference
in behavior of "short" and "long" JJ's. Length scales somewhat
different from that in \cite{Clem2} were derived in Ref.
\cite{Koshel}, based on simulations for $N=15$. In general,
different conclusions about the fluxon size could be made from
numerical simulations for different $N$: short scale $\sim \lambda
_J$ for $N$=7 \cite{SBP}, and $N$=19 \cite{Klein1}; and long scale
$\sim \lambda _c$ for $N$=50 \cite{Shaf}. Long scale, $\sim
\lambda _c$, variation of $B$ for $N>1000$, was deduced from
experimental observations of the fluxon in HTSC \cite{Kirtley}.
From the analysis above it is clear that such "discrepancy" is a
consequence of strong $N-$ dependence of the fluxon shape,
however, $\lambda _{J(eff)} = \lambda(0)$ is always $\simeq
\lambda_J$, as shown in inset of Fig. 2 b).

\subsection{Flux-flow characteristics}

\ \ \ \ \ Variation of fluxon shape in dynamic state is
particularly interesting both because it has a direct impact on
$c-$axis transport properties and because fluxon shape becomes
very unusual at high propagation velocities, see Fig. 1. When bias
current is applied, a fluxon starts moving along the junction with
a velocity determined by the balance between Lorentz and viscous
friction forces. This results in appearance of a flux-flow branch
in current-voltage characteristics (IVC's) with average DC voltage
$V_{i_0}/V_{0}=u/c_{0}$, ($V_{0}=\frac{\hbar \pi c_{0}}{2eL}$) in
the JJ containing the fluxon and zero for $i\neq i_0$. Therefore,
current-velocity characteristics are equivalent to IVC's. In a
single JJ, fluxon is a relativistic object. It gets Lorentz
contracted \cite{Laub} when the fluxon velocity approaches Swihart
velocity, i.e. the phase velocity of electromagnetic waves in the
transmission line formed by the junction. However, in SJJ's
according to the multi-component solution, only $F_1$ component of
the fluxon gets Lorentz contracted at the lowest velocity $c_1$.
The weight of this component decreases with $N$, i.e., there is
less contraction for larger $N$.

\begin{figure}
\begin{center}
\includegraphics[width=8cm]{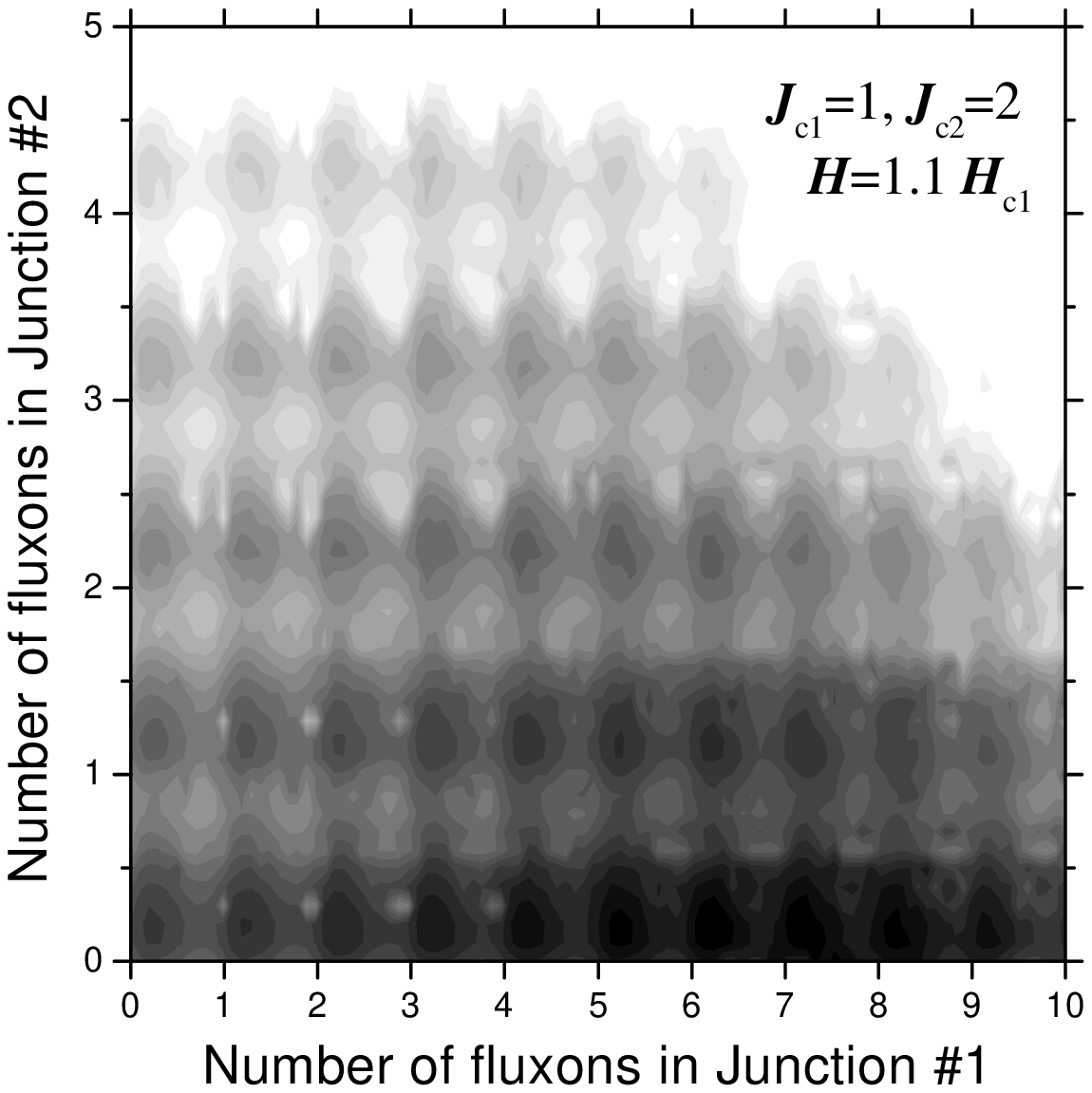}
\caption{Gray scale plot of Gibbs free energy versus the number of
fluxons in double SJJ's at $H=1.1H_{c1}$. Darker regions
correspond to smaller Gibbs energy. The existence of multiple
quasi-equilibrium states is seen. From Ref. \cite{Modes}.}
\label{Gibbs}
\end{center}
\end{figure}

So, what happens with the fluxon
in SJJ's: does it gets contracted at $u=c_1$ or not? Systematic
analysis of the double SJJ's showed that the answer
strongly depends on parameters of SJJ's \cite{FluxonD}. From Fig.
1 it is seen that the fluxon in identical double SJJ's is well
described by the multi-component solution up to $u=c_1$. However,
in general, partial linearization procedure fail at $u=c_1$
because $\varphi_{i \ne i_0}$ is no longer small. It was found
\cite{FluxonD}, that for $\frac{J_{c2}\Lambda _2}{J_{c1}\Lambda
_1}>1$ and $\kappa _2>1$, a "single component" ($F_2$) solution
\cite{Modes} is realized in double SJJ's. This solution was
anticipated in Ref. \cite{Modes} and is characterized by
relationship $sin(\varphi_1) \simeq \kappa_2 sin(\varphi_2)$.
Note, that it becomes exact at $u=c_1$ for the above mentioned
parameters. Apparently, the single $F_2$ component fluxon does not
contract at $u=c_1$.

In Fig. 4, numerically simulated IVC's are shown for a single
junction $N=1$, and SJJ's with $N=$2, 3, 5 and 21. Simulations
were done for annular junction geometry with $L=100\lambda_J$,
parameters typical for Bi2212 HTSC and damping coefficient
$\alpha_i=0.05$. Vertical dotted lines represent lowest
characteristic velocities: $c_1(N=2)\simeq 0.8165$,
$c_1(N=3)\simeq 0.7653$ and $c_1(N=21)\simeq 0.7071 c_0$. Clear
velocity matching step at $u\rightarrow c_0$ is seen for a single
junction \cite{Scott}. Vertical portions of IVC's at $u\rightarrow
c_1$ are also seen for $N=$ 2 and 3, as a result of partial
Lorentz contraction of the fluxon, see Fig. 1. However, for $N=5$
and 21 there is no peculiarity in IVC's at a velocity matching
condition $u=c_1$, indicating absence of fluxon contraction. In
Ref. \cite{FluxoN} it was found out that Lorentz contraction at
$u=c_1$ is absent for SJJ's with $N>3$. Note that there are very
little changes in IVC's for $N\geq 5$ (the IVC's for $N=5$ and 21
are almost indistinguishable in Fig. 4). This is due to the fact
that dissipation is mostly determined by the fluxon core region,
which remains almost unchanged for $N\geq 5$, see $N-$dependence
of $\lambda(0)$ in inset of Fig. 2 b).

\subsection{Cherenkov radiation}

\ \ \ \ \ Due of the lack of Lorentz contraction, the fluxon can
propagate with a velocity larger than $c_1$ as independently
proposed in Refs. \cite{HechtfPRL,Modes}. Such super-relativistic
motion is accompanied by "Cherenkov" radiation
\cite{Malomed,HechtfPRL,FluxonD,Kleiner62,FluxoN,Goldob}. An
important insight to Cherenkov radiation can be obtained from the
multi-component solution. For $u>c_n$, characteristic lengths
$\tilde \lambda_{m<n}$ become imaginary, characteristic for "wave"
solutions. Indeed, far from the fluxon center we can further
linearize Eq.(8), and obtain that the solution for $F_{m<n}$ is a
traveling wave with the dispersion relation \cite{Modes,FluxoN}:

\begin{figure}
\begin{center}
\includegraphics[width=8cm]{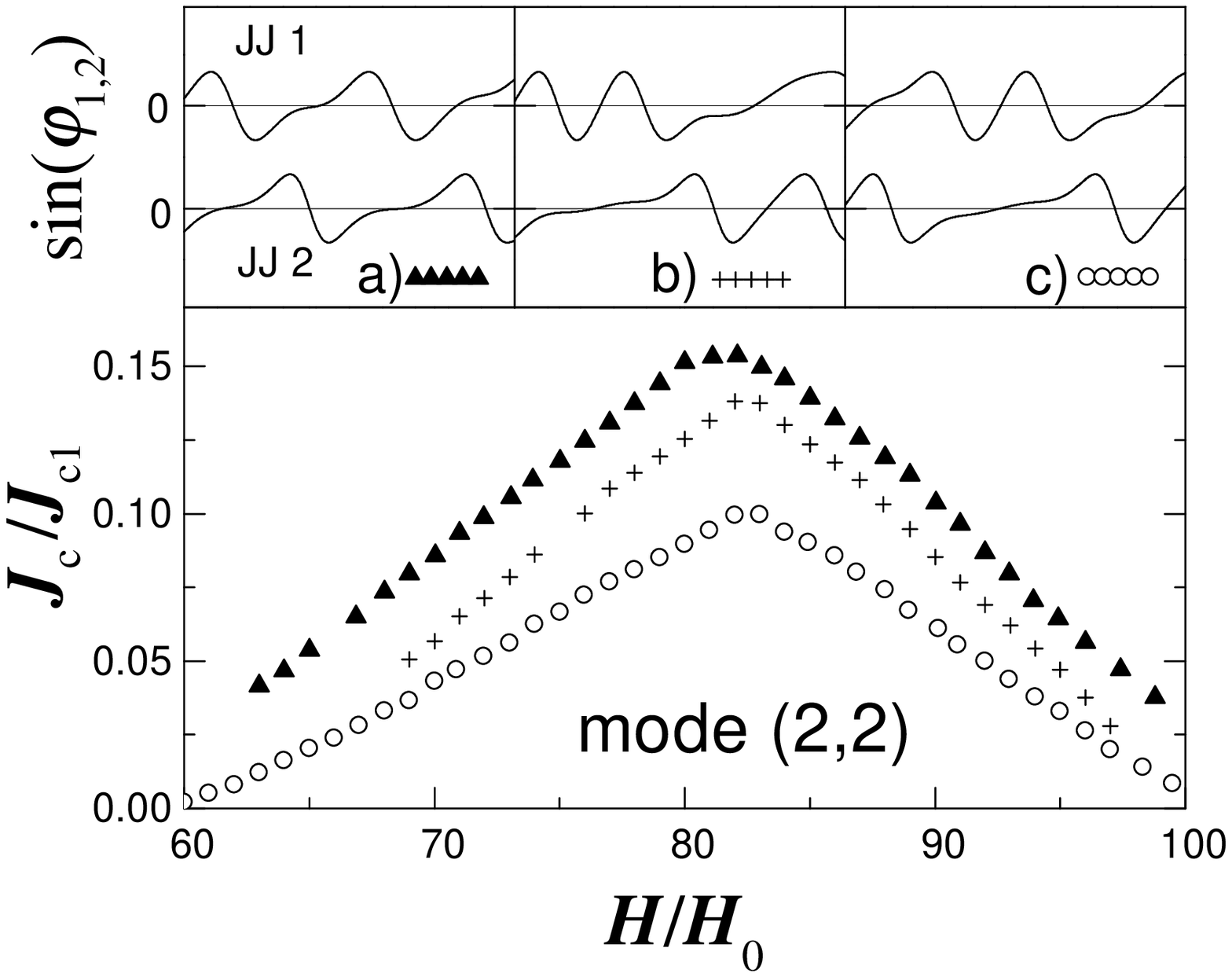}
\caption{Three simulated $I_c(H)$ sub-branches of double SJJ's,
corresponding to submodes of the (2,2) mode with different fluxon
sequence. Insets a-c) show spatial current distributions in
junctions 1 and 2 for each submode at $H/H_0=80$. From Ref.
\cite{SubModes}} \label{subm}
\end{center}
\end{figure}

\begin{equation}
\kappa_m^{-2}=u^2/\tilde\omega_p^2-\lambda_m^2,\ \omega=u\kappa_m,
%\eqnum{18}
\label{Eq.18}
\end{equation}

\noindent where $\tilde\omega_p^2=\omega_p^2\sqrt{1-J_b^2}$. Those
are Josephson plasma waves \cite{Klein1,SakPed} moving along with
the fluxon at velocity $u$ \cite{Modes}.

In Fig. 5, numerically simulated phase distributions for a rapidly
moving fluxon ($c_6>u\simeq 0.928 c_0>c_5$) in $N=11$ SJJ's are
shown for a central junction, containing the fluxon, $i_0=6$ (top
panel); $i=4$ (middle panel); and the outmost junction $i=1$
(bottom panel). Parameters of SJJ's are the same as in Fig. 4. The
six lowest characteristic velocities for $N=11$ are $c_1=0.7132
c_0$, $c_2=0.7320 c_0$, $c_3=0.7653 c_0$, $c_4=0.8165 c_0$,
$c_5=0.8913 c_0$, $c_6= c_0$. Well defined oscillations are seen
behind the fluxon (the fluxon is moving from right to left). The
spectra of oscillations are shown in insets a) for $i=6$ and b)
for $i=4$ (solid line) and $i=1$ (dotted line). Clear beatings of
oscillations are seen in JJ's 6 and 1, which is a
signature of coexistence of multiple wave lengths. Indeed, spectra
of oscillations exhibit several maxima corresponding to odd plasma
modes, as indicated by arrows in insets a and b).

Numerical simulations support the
assumption \cite{Modes} that Cherenkov radiation at
$u>c_n$ is due to degeneration of fluxon components $F_{m<n}$ into
Josephson plasma waves \cite{FluxonD,FluxoN}. Namely, it was observed that:
(i) Plasma mode $m$ is generated when $u>c_m$. (ii) Only those
modes which constitute the fluxon, appear in Cherenkov radiation.
E.g., for a fluxon in the central junction of a stack with odd $N$
there are no even modes, see Eq. (14) and Fig. 5. (iii) Relative
amplitudes of oscillations are related to weights coefficients of
the corresponding fluxon component. For the case of Fig. 5, weight
coefficient of mode 3 in junction 4, $K^{-1}_{4,3}=0$, and plasma
mode 3 is absent in the spectrum. On the contrary, for the outmost
junction, the weight coefficient of mode 3, $K^{-1}_{1,3} \simeq
0.118$, is significantly larger than for mode 1, $K^{-1}_{1,1}
\simeq -0.043$, and the maximum of amplitude corresponds to plasma
mode 3, see inset b) in Fig. 5.

\section{MULTI-FLUXON CONFIGURATIONS}

\ \ \ \ \ Due to repulsive fluxon interaction, fluxons are
believed to form a regular fluxon lattice (FL) at large parallel
magnetic field. Since magnetic induction of a fluxon is spread
over many layers, see Fig. 3, it is not favorable to have a
perfect translational symmetry along the $c-$axis, and FL is
expected to be rhombic \cite{Levit}. However, the perfect FL can
be achieved only when fluxons strongly interact with each other,
i.e., when spacing between fluxons along the $ab$-plane is $\sim
\lambda _J$. For Bi and Tl based HTSC, the corresponding magnetic
field, $\Phi _0/\lambda_J s \sim$ 1T is three orders of magnitude
larger than the lower critical field $H_{c1}^{\mid \mid }\sim $mT
\cite{KrLaRy,Nakamura}.

\begin{figure}
\begin{center}
\includegraphics[width=14cm]{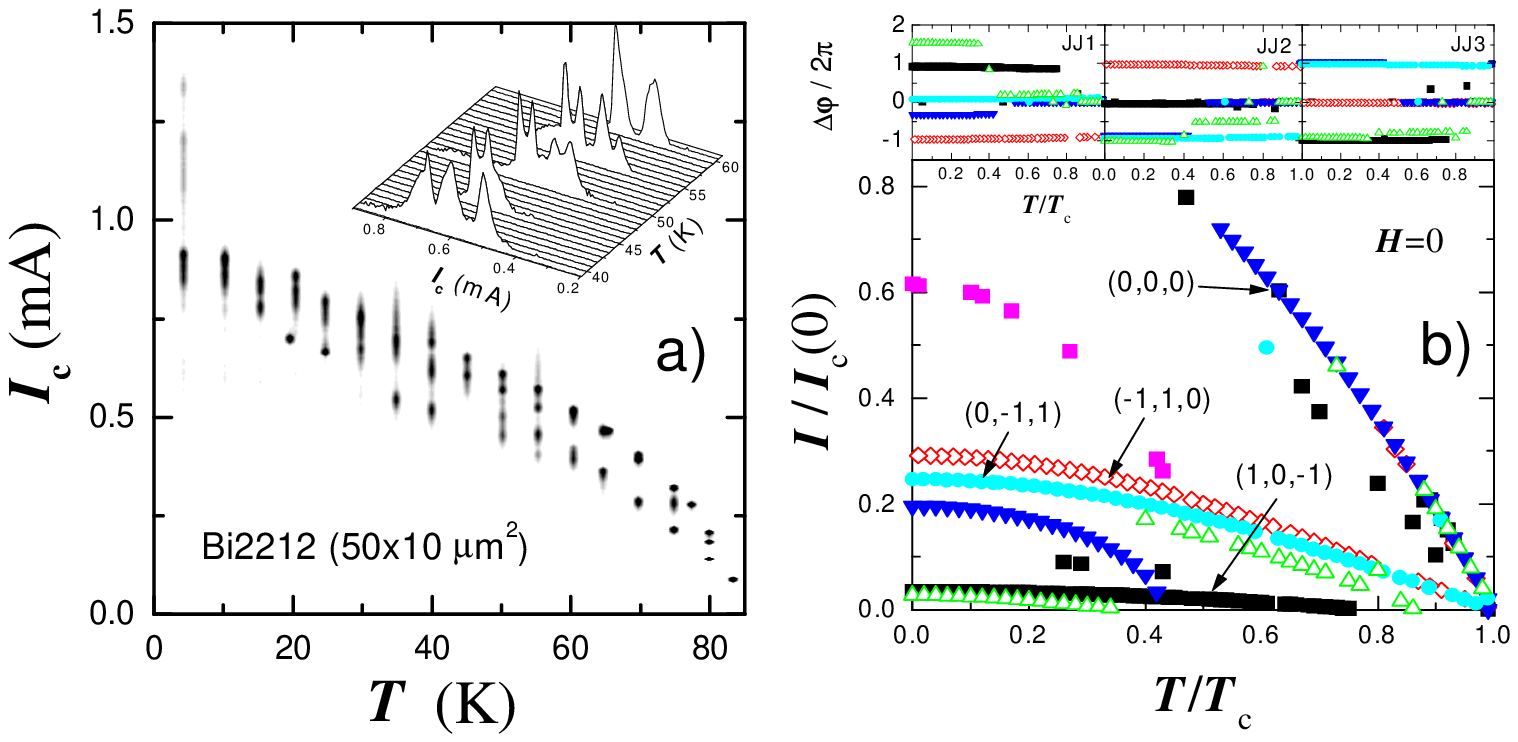}
\caption{a) A gray scale plot of the probability distribution of
$I_c(T)$ for a Bi2212 mesa. Inset demonstrates multiple-peak
structure of $P\left( I_c\right)$ for several $T$. b) Simulated
$I_c(T)$ for three SJJ's with Bi2212 parameters at $H$=0. Multiple
$I_c(T)$ branches are clearly seen. Branches corresponding to
(0,0,0) and simplest fluxon-antifluxon modes are marked. Insets
show distribution of fluxons in the stack. From Ref.
\cite{Compar}} \label{IcT}
\end{center}
\end{figure}

Fluxon distribution in layered superconductors at low fields,
$H_{c1}^{\mid \mid }<H<\Phi _0/\lambda_J s$, is still a matter of
controversy. Levitov \cite{Levit} have found ''phyllotaxis'' and
bifurcations in the FL at low fields, due to existence of multiple
metastable FL's with approximately equal energies. In Ref.
\cite{Watson}, this approach was extended using a ''growth''
algorithm, allowing variation of the FL along the $c$-axis.
Although restricted to constant space periodicity along layers,
the model showed the existence of a large variety of
quasi-periodic or aperiodic fluxon configurations in the $c$-axis
direction. At low fields, the fluxon system becomes completely
frustrated and tends to be chaotic \cite{Watson}. Recent numerical
simulations revealed that fluxons in SJJ's may form buckled chains
\cite{Shaf,Hu} instead of the regular FL. Clearly, the lattice
description of multi-fluxon distribution becomes unappropriate at
low magnetic fields.

\subsection{Quasi-equilibrium fluxon modes and submodes}

\ \ \ \ \ Recently it was shown that multiple quasi-equilibrium
fluxon configurations (modes) exist in long, $L\gg \lambda _J$,
strongly coupled SJJ's \cite {Modes}. Fig. 6 shows a numerically
simulated gray scale plot of Gibbs free energy versus the number
of fluxons in double SJJ's for applied parallel field slighltly
above the lower critical field, $H = 1.1 H_{c1}$. Parameters of
the stack are $J_{c2}=2 J_{c1}$, $\lambda_{Si}=0.1\lambda_{J1}$,
$t_i=d_i=0.01\lambda_{J1}$, $L=50 \lambda_{J1}$. Darker regions
correspond to smaller Gibbs energy. The absolute minimum,
representing a thermodynamic equilibrium, correspond to 7 fluxons
in junction 1 and zero fluxons in junction 2. However, it is seen
that there is a number of other fluxon configurations (modes) for
which local minima of Gibbs free energy is achieved. All those
modes are stable and represent quasi-equilibrium states of the
stack. The modes are characterized by different fluxon
distributions in the stack and do not have $a-priori$ any
translational symmetry. We will use a string $(n_1,...n_i,...n_N)$
as a short notation of fluxon modes, where $n_i$ is a number of
fluxons in junction $i$ and $N$ is the number of junctions in the
stack. E.g., the equilibrium state in Fig. 6 corresponds to (7,0)
mode.

In Ref. \cite{SubModes} existence of fluxon submodes was
demonstrated, which have the same number of fluxons but are
different with respect to fluxon sequences and symmetry of phase
differences in the junctions (along the planes). Insets a-c) in
Fig. 7, show three different submodes of (2,2) mode for a double
stack with the same parameters as in Fig. 6. Those submodes have
nearly equal self energies, i.e. nearly equal probability,
however, correspond to distinctly different $I_c(H)$ sub-branches.

\begin{figure}
\begin{center}
\includegraphics[width=14cm]{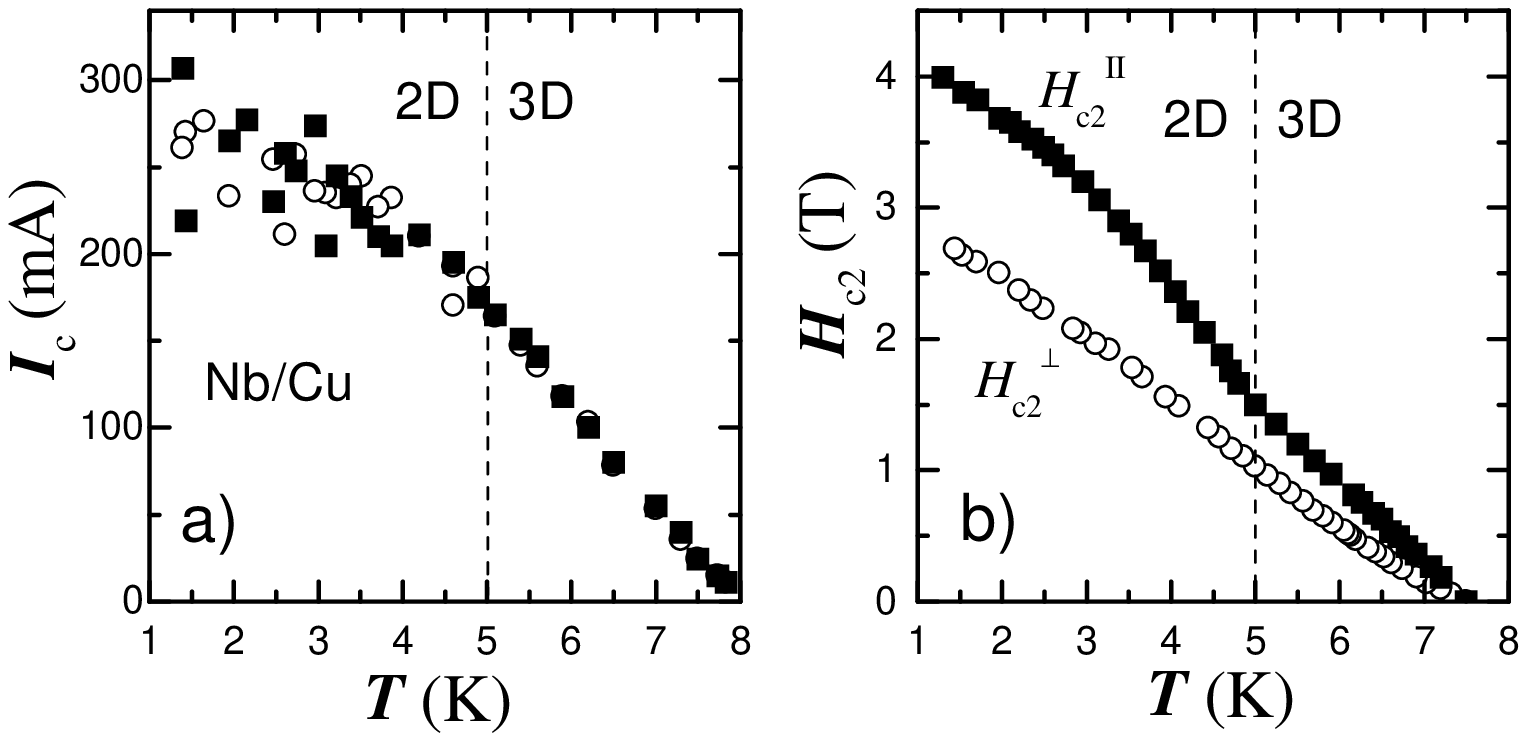}
\caption{a) $I_c\left( T\right)$, for Nb/Cu multilayer. Different
symbols correspond to different runs. It is seen that strong
fluctuations of $I_c$ take place in the 2D state. b) The upper
critical field parallel (solid rectangles) and perpendicular (open
circles) to layers. A kink in $H_{c2}^{\mid \mid }$ represents the
3D-2D crossover. Data from Ref. \cite{Compar}} \label{IcT}
\end{center}
\end{figure}

The number of fluxon modes and submodes increases rapidly with
increasing number of junctions and fluxons. From mathematical
statistics it follows that for a stack with N non-identical
junctions with M fluxons, the total number of fluxon modes
(indistinguishable fluxon sequences) is

\begin{equation}
N_{modes}=\frac{(N+M-1)!}{(N-1)!M!}, \label{modes}
\end{equation}

\noindent and the number of submodes (distinguishable fluxon
sequences) is

\begin{equation}
N_{submodes}=N^M. \label{Submodes}
\end{equation}

A specific feature of SJJ's is existence of stable
fluxon-antifluxon pairs \cite{Compar}. Due to attractive
interaction between fluxons and antifluxons, they tend to collapse
both in a single junction or type-II superconductor. However, in
SJJ's fluxons can not move across S-layers. Therefore, a fluxon
and an antifluxon in neighbor junctions do not collapse, but form
a bound pair. Furthermore, once in the stack, such pair can not be
removed (unless destroyed) by transport current because there is
no net Lorentz force on the pair. Fluxon-antifluxon pairs are
likely to be present in SJJ's at $H<H_{c1}$ and can dominate
transport properties at zero field \cite{Compar,Kleiner62}.

Due to existence of multiple quasi-equilibrium fluxon (and
fluxon-antifluxon) modes/ submodes, the state of SJJ's is not
well defined. It can be described only statistically with a
certain probability of being in any of the modes. Neither the
number of fluxons in each junction nor even the total number of
fluxons in the stack is fixed for given $H$ and $T$. Moreover,
since different fluxon modes/submodes may have similar energies,
the system is frustrated and exhibit strong fluctuations. This may
account for unusual $c$-axis transport properties of both HTSC
mesas \cite{Mros,SubModes,Compar} and low-$T_c$ SJJ's
\cite{NbCuH,Compar}.

\subsection{Multiple-valued critical current}

\ \ \ \ \ Statistical analysis is crucial for probing  multiple
quasi-equilibrium states in SJJ's. Fig. 8 a)
presents in a gray scale a probability distribution of critical
current $P\left( I_c\right)$ vs. $T$ for 50 $\mu $m long Bi2212
mesa with $N=5$ intrinsic SJJ's, at earth magnetic field. Darker
regions correspond to a larger probability. To obtain $P\left(
I_c\right) $, 5120 switching events from zero-voltage state were
measured at each $T$ \cite{Mros,Compar}. Inset in Fig. 8 a) shows
$P\left( I_c\right) $ for 40 K $<T<$ 60 K. It is seen that
$P\left( I_c\right)$ is very wide and consists of several
superimposed peaks, attributed to different fluxon modes
\cite{Mros,Compar}.

\begin{figure}
\begin{center}
\includegraphics[width=14cm]{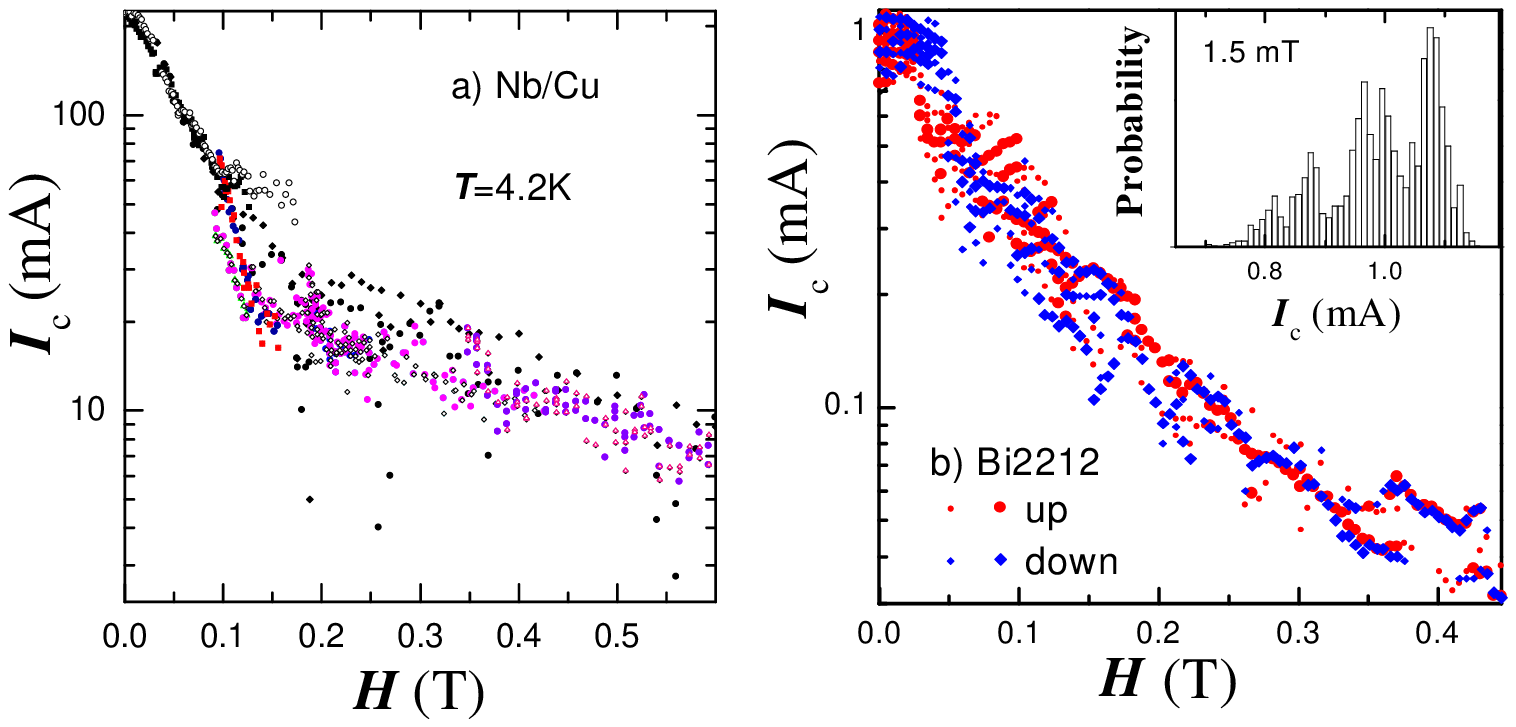}
\caption{$I_c(H)$ dependencies for a) Nb/Cu multilayer, and b)
Bi2212 mesa. Symbols correspond to maxima in $P(I_c)$, shown in
inset. Data from Ref. \cite{Compar}} \label{Fig5m}
\end{center}
\end{figure}

In Fig. 8 b) simulated $I_c(T)$ dependencies are shown for $N=3$
long ($L = 10 \lambda_J$) SJJ's with Bi2212 parameters, at $H$=0.
Different symbols represent separate runs with
increasing or decreasing $T$ and/or different initial
conditions. It is seen that $I_c(T)$ consists of multiple distinct
branches, in qualitative agreement with experiment. Insets in Fig.
8 b) show amount of fluxons in the junctions. It can be seen that
each $I_c(T)$ branch corresponds to a particular fluxon or
fluxon-antifluxon mode. I want to emphasize that $I_c$ is multiple
valued even at $H=0$. This is due to stability of
fluxon-antifluxon pairs in SJJ's as discussed above. Branches
corresponding to simplest fluxon-antifluxon modes (-1,1,0),
(0,-1,1) and (1,0,-1) are marked in Fig. 8 b).

Fig. 9 a) shows $I_c\left(
T\right) $ for a Nb/Cu (20/15 nm $N=$ 10) multilayer. Different
symbols correspond to different runs. Measurements were done at
$H\sim $ 10 mT along layers. It is seen that $I_c$ strongly
fluctuates below $\sim $5 K. Fig. 9 b) shows the upper critical
field $H_{c2}(T)$ parallel (solid rectangles) and perpendicular
(open circles) to layers. A kink in $H_{c2}^{\mid \mid }(T)$
reflects dimensional 3D-2D crossover, which is well studied in
those systems [48,53,63-66]. The 3D-2D crossover is one of
peculiar properties of superconducting multilayers
\cite{Takahashi}. In the 3D state a multilayer behaves as a bulk
superconductor, while in the 2D state - as a stack of Josephson
junctions \cite{NbCuJ,NbCuH}. Dashed lines in Fig. 9 show, that
strong fluctuations of $I_c$ appear in the 2D state reflecting
appearance of SJJ's. This unambiguously demonstrates that multiple
valued critical current is a genuine feature of SJJ's, due to
existence of multiple fluxon modes.

\subsection{Magnetic field dependence of the critical current}

\ \ \ \ \ The dependence of $I_c$ on in-plane magnetic field is a
crucial test for dc Josephson effect. In a single JJ, Fraunhofer
modulation of $I_c(H)$ occurs as a result of flux quantization
\cite{Barone}. For SJJ's it was predicted that the periodicity of
$I_c(H)$ is \cite{Bul92}

\begin{equation}
\Delta H=\Phi _0/Ls.  \label{Eq.6}
\end{equation}

\noindent However, experimental $I_c(H)$ both for HTSC
\cite{Kleiner94,LatyshevPRL,SubModes} and low-$T_c$
\cite{Amin,Nevirk,KleinSak,NbCuH} SJJ's do not show such
periodicity. So, what's wrong with dc-Josephson effect in SJJ's?

Experimental $I_c(H)$ for Nb/Cu multilayer ($N$=10, $L=$20 $\mu$
m) and Bi2212 mesa ($N=$5, $L=$ 20 $\mu$m) are shown in Fig. 10 a)
and b), respectively, at $T=$ 4.2 K. Note a logarithmic scale of
$I_c$. In Fig. 10 a), different symbols represent different zero-
field-cooled runs with increasing field. Fig. 10 b) shows a result
of statistical analysis. Large and small symbols in Fig. 10 b)
represent main and secondary peaks in $P(I_c)$, respectively, for
increasing (circles) and decreasing (diamonds) field. The
probability distribution with several maxima is shown in inset. It
is seen that $I_c$ is multiple valued and, even though one could
probably recognize certain branches, there is no periodic
Fraunhofer modulation.

\begin{figure}
\begin{center}
\includegraphics[width=14cm]{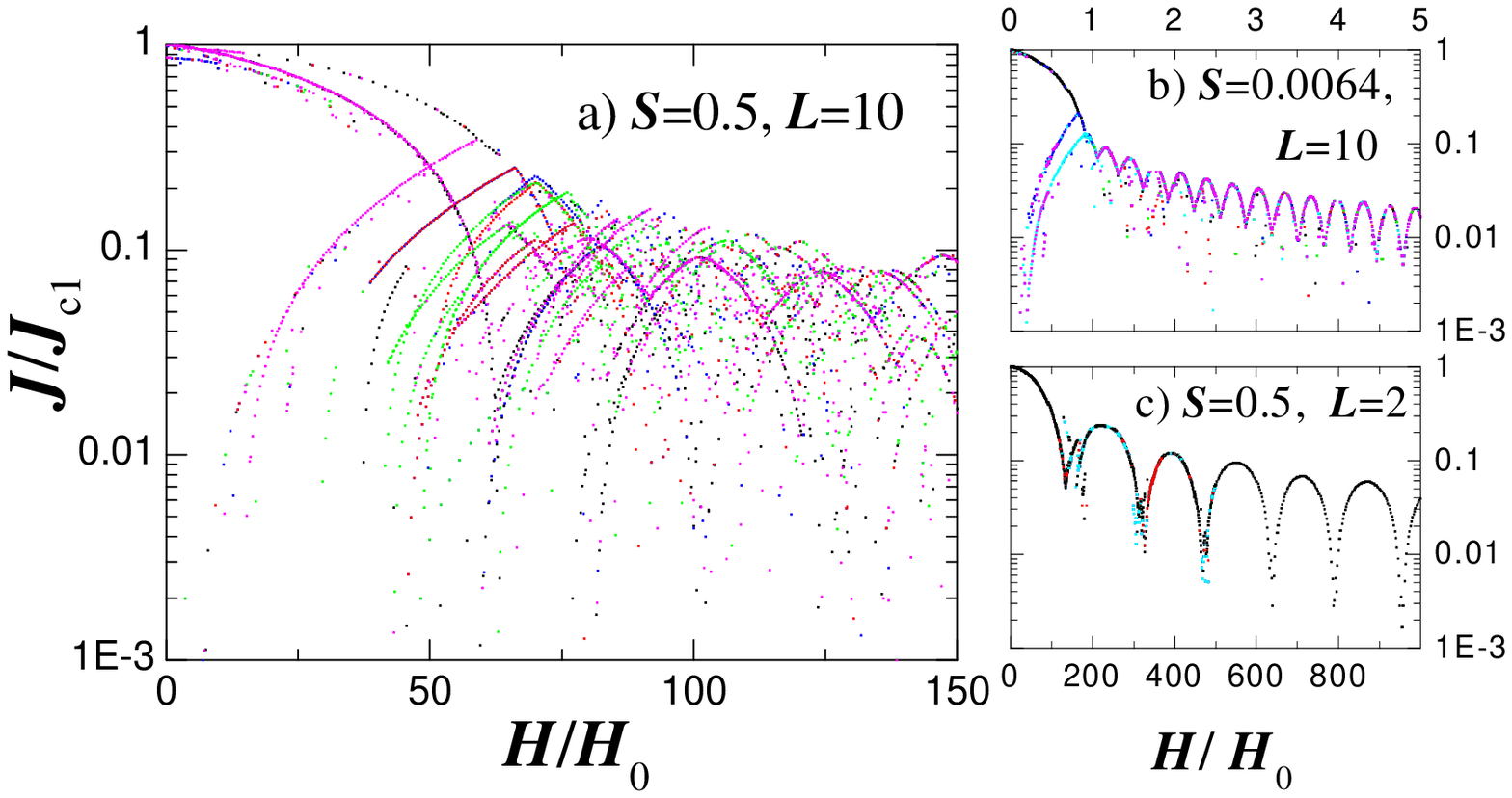}
\caption{Simulated $I_c(H)$ for a) long, strongly coupled b) long,
weakly coupled and c) short, strongly coupled double SJJ's. It is
seen that Fraunhofer modulation is obscured in long, strongly
coupled SJJ's due to existence of multiple quasiequilibrium fluxon
modes. However, periodicity of $I_c(H)$ is restored either in
weakly coupled or short SJJ's. Data from Ref. \cite{SubModes}}
\label{IcHn}
\end{center}
\end{figure}

Fig. 11 presents numerically simulated $I_c(H)$ patterns for a)
long strongly coupled ($L=10\lambda_{J1}$, $S=0.495$, $d/\lambda_S
=0.1$); b) long weakly coupled ($L=10\lambda_{J1}$, $S=0.0064$,
$d/\lambda_S =5$) and c) short strongly coupled
($L=2\lambda_{J1}$, $S=0.495$, $d/\lambda_S =0.1$) double SJJ's
with $J_{c2}=2J_{c1}$. Note logarithmic scale for $I_c$. Plots
contain several runs for increasing or decreasing field with
different initial conditions. From Fig. 11 a) it is seen that the
$I_c(H)$ pattern for long, strongly coupled SJJ's consists of
multiple closely spaced branches. Each branch is characterized by
a certain fluxon distribution, i.e., corresponds to a certain
fluxon mode or submodes, see Fig. 7. The $I_c(H)$ pattern does not
exhibit clear periodicity because $I_c$ switches randomly between
closely spaced branches. This does not mean that something is
wrong with dc Josephson effect, but rather is a consequence of
magnetic coupling between junctions.

From Figs. 11 b) and c) it is seen that periodicity of $I_c(H)$ is
restored either when the coupling is decreased or a stack becomes
short compared to $\lambda_J$. For weakly coupled SJJ's, $I_c(H)$
is simply determined by the smallest critical current of
individual JJ's, while in short SJJ's there are no fluxons and,
consequently, no fluxon modes. A transition from aperiodic to
periodic modulation of $I_c(H)$ with decreasing $L/\lambda_J$ was
observed both for Nb/Cu multilayers \cite{NbCuH} and for
YBa$_2$Cu$_3$O$_{7-x}$ \cite{Ling} close to $T_c$.

\subsection{Multiple flux-flow sub-branches}

\ \ \ \ \ In sec. 3.4 we considered viscous motion of a single
fluxon. For a multi-fluxon configuration, time averaged flux-flow
voltage is proportional to the number of fluxons in SJJ's and a
fluxon velocity. For large $H$, the ratio $V/H$ is:

\begin{equation}
V/H \simeq Nsu.  \label{Eq.23}
\end{equation}

\noindent Note that it is independent on the junction length. A
characteristic feature of SJJ's is splitting of Swihart
velocities, see Eq.(17), and corresponding splitting of flux-flow
IVC's \cite{SBP,Klein1,SakPed}. Flux-flow phenomenon was
intensively studied both for HTSC [30,37-41,74,75] and low-$T_c$
\cite{Ust93,Monaco,Thyssen} SJJ's due to possible application as a
source of microwave radiation \cite{Machida}. From application
point of view, the most interesting is coherent (in-phase) fluxon
propagation \cite {SBP} at the highest characteristic velocity
$c_N$, because it could generate narrow linewidth radiation with
large amplitude \cite{Machida}. However, so far the measured
radiation linewidth from SJJ's was much wider than expected
\cite{Shitov,HechtfPRL}.

In Fig. 12 a) flux-flow IVC's of long ($L= 50 \mu$m, $N=5$) Bi2212 mesa
are shown for $H = 0.148 - 0.396$ T
with a step $\simeq$ 0.025 T, at $T$=4.2 K. Magnetic field was
applied along the $ab$-plane and perpendicular to the longest side
of the mesa. It is seen, that a low resistance branch develops in
IVC's with increasing $H$. The 20 $\mu $m long mesa on the same
single crystal showed the same $V/H$ value, in agreement with Eq.
(23). However, IVC's were more complicated due to presence of
Fiske steps \cite{Fiske}. The maximum flux-flow velocity is close
to the lowest characteristic velocity $c_1 \simeq 2.5-3\times
10^{7} cm/s$. This value is consistent with observations by
different groups [37-41,74,75] and is in agreement with the Fiske
step voltage observed in Bi2212 mesas \cite{Fiske}.

\begin{figure}
\begin{center}
\includegraphics[width=14cm]{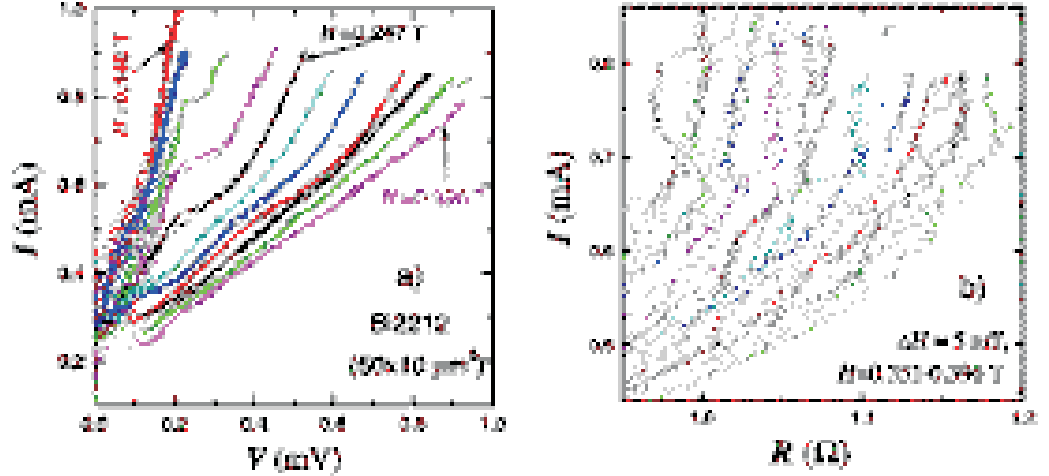} \caption{a) Flux-flow IVC's of 50 $\mu$m long Bi2212 mesa
with $N=5$ intrinsic SJJ's at $T$=4.2 K are shown for $H$ from
0.148 to 0.396 T with a step $\simeq$0.025 T. b) Enlarged top
parts of $I$ vs. $R=V/I$ curves are shown for $H$ varying with a small step $\simeq$5 mT in the range $H$%
=0.351$\div $0.396 T, at $T$=4.2 K. Closely spaced flux-flow
sub-branches are seen. Data from Ref. \cite{Compar}} \label{BiFF}
\end{center}
\end{figure}

From Fig. 12 a) it is seen that flux-flow branches exhibit strong
fluctuations. At small voltages, the spread in IVC's reflects
fluctuations in $I_c$, as discussed in sec. 4.2. In Fig. 12 b)
enlarged top parts of $I$ vs. $R=V/I$ curves for the same mesa are
shown at magnetic fields in the range 0.351 - 0.396 T, varying in
small steps $\simeq$ 5 mT, at $T$=4.2 K. The characteristic
feature of IVC's from Fig. 12 is that the curves are not just
uniformly wide, but rather consist of multiple closely spaced but
distinct sub-branches. The IVC's switch hysteretically between
sub-branches when current is swept back and forth. Varying $H$, we
only change the set of available sub-branches but do not change
the shape of a particular sub-branch.

Similar multiple flux-flow sub-branches were observed for Nb/Cu
multilayers \cite{NbCuH} and HTSC mesas \cite{Lee2,Doh}. In the
latter two papers it was suggested that sub-branching is due to
splitting of Swihart velocities. Indeed such splitting was
observed for low-$T_c$ SJJ's \cite{Ust93,Monaco,Thyssen} and
possibly for HTSC mesas \cite{KrIEEE}. In those cases branches
revealed {\it different} $V/H$ scaling, corresponding to different
$c_n$, see Eq.(23). On the contrary, tiny sub-branches shown in
Fig.12 b) (as well as those observed in Refs. \cite{Lee2,Doh})
have approximately the same $V/H$ scaling corresponding to the
lowest velocity $c_1$. Furthermore, because of small $N$=5,
splitting of $c_n$ should correspond to almost two orders of
magnitude larger splitting than that in Fig. 12 b) \cite{KrIEEE}.
It should also be noted that the number of sub-branches is not
limited by the number of SJJ's in the mesa. All this rules out
explanation of multiple flux-flow sub-branches, shown in Fig. 12
b), in terms of splitting of Swihart velocities.

In Refs. \cite{NbCuH} it was proposed that multiple flux-flow
sub-branches correspond to propagation of different
quasi-equilibrium fluxon modes. This was supported by direct
numerical simulations \cite{PhysB,Compar,Kleiner62}. Fig. 13
presents numerical modeling of IVC's for the mesa from Fig. 12 (
$N=5$, $J_c \sim 10^3 A/cm^2$, $s=15 \AA$, $d=3 \AA$, $\lambda_J
\simeq 1 \mu m$, $L/\lambda_J =50$ and damping parameters
$\alpha_i = 0.05$). Junctions are identical, except for larger
thickness of the bottom electrode, $d_1=6 \AA$, in order to
imitate the bulk crystal beneath the mesa. Two successive
back-and-forth current sweeps are shown (denoted as A and B) at
in-plane magnetic field $H=10^4 H_0 \simeq 2000 $ Oe. Top three
panels show overall IVC's of the stack, $V=\sum V_i$, with
successive "zoom-in" to a small voltage region: panel a)
demonstrates IVC's in a large scale, panel b) presents the
flux-flow branch, and panel c) shows details of switching between
supercurrent ($V=0$) and flux-flow branches. In the bottom row,
the left and right groups of plots represent data for sweeps A and
B, respectively. In each group, the left plots represent
individual IVC's of all five junctions (successively shifted along
$V-$ axis for clarity), while the right plots show amount of
fluxons in each junction, determined as
$[\varphi_i(L)-\varphi_i(0)]/2\pi$ (lines), and the total amount
of fluxons in the stack (symbols). Despite nominally the same
external conditions, the IVC's A and B are apparently different.
For the IVC-A only three ($i=$2, 3, and 4), while for the IVC-B
all five junctions switch to the normal branch (shown by dotted
lines in Fig. a)) at the end of the flux-flow branch.

\begin{figure}
\begin{center}
\includegraphics[width=13cm]{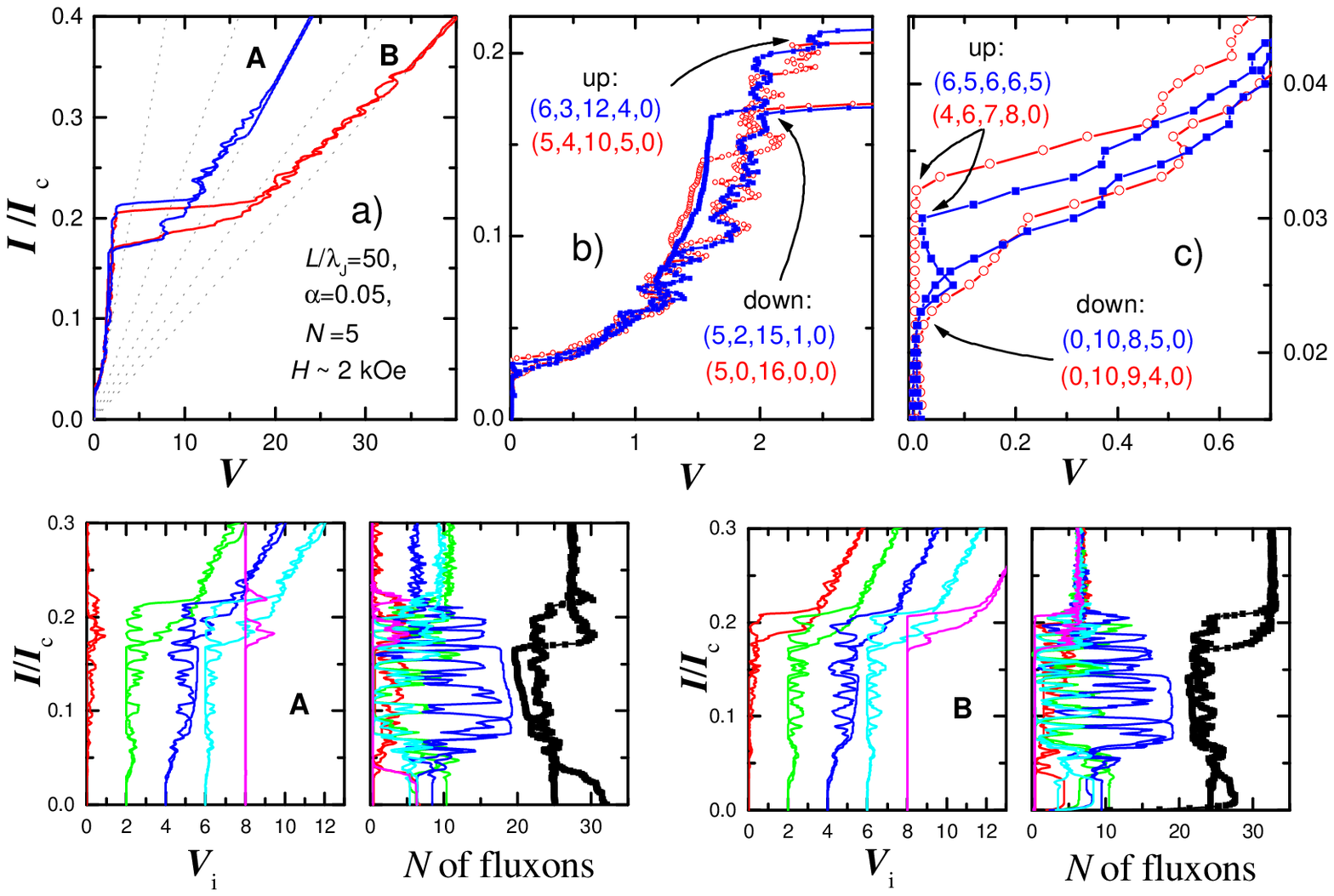}
\caption{ Simulated IVC's are shown for two back-and-forth current
sweeps (denoted as A and B) for the Bi2212 mesa from Fig. 12. Top
three panels show overall IVC's $V=\sum V_i$, with successive
"zoom-in" to a small voltage region from panel a) to c). Fluxon
modes realized at the end and the beginning of flux-flow branches
for both sweeps are specified. In the bottom row, IVC's of
individual junctions (successively offset along the $V-$ axis for
clarity) and the number of fluxons in the junctions are shown. It
is seen that fluctuation of the flux-flow voltage is due to
switching between fluxon modes. } \label{IVC_num}
\end{center}
\end{figure}

From Fig. 13 b) it is seen that flux-flow branches exhibit strong
fluctuations, caused by variation of fluxon numbers in junctions,
as shown in the bottom row of Fig.13. Each time fluxons leave the
stack on one side, other fluxons entering the stack on the
opposite side have freedom to choose between all possible
quasi-equilibrium fluxon modes. Each fluxon mode is represented by
a distinct flux-flow sub-branch, with slightly different $I_c$ and
flux-flow voltage. Two of such sub-branches can be seen at $I/I_c
= 0.1 - 0.15$ in Fig. 13 b). Fluxon modes realized at the end and
the beginning of flux-flow branches for both sweeps are specified
in Figs. 13 b) and c), respectively. Note that in both cases the
maximum fluxon velocity is close to the lowest characteristic
velocity $\sim c_1$, i.e., splitting of $c_n$ is not the cause of
sub-branching of IVC's. Fig. 13 c) demonstrates that critical
currents and return currents are also different for the two
sweeps, i.e., $I_c$ is multiple valued. As discussed in sec. 4.2.
different $I_c$'s correspond to different modes, indicated in Fig.
13 c).

Fig. 13 reproduces main features of experimental IVC's from Fig.
12. In Ref. \cite{Kleiner62} it was shown that tiny sub-branches
due to fluxon-antifluxon modes exist also at zero field. Switching
between fluxon modes/submodes and correlated fluctuations of
flux-flow voltage would cause extremely wide radiation line width,
consistent with numerical simulations \cite{Machida} and
experimental radiation detection from HTSC SJJ's \cite{HechtfPRL}.

\section{CONCLUSIONS}

\ \ \ \ \ In conclusion, single and multi-fluxon
properties of layered superconductors and stacked Josephson
junctions were analyzed. We have seen that behavior of SJJ's can be
qualitatively different from that of single Josephson junctions or
type-II superconductors. To a large extent, anomalous properties
are due to unusual fluxon shape and existence of multiple
quasi-equilibrium fluxon modes in long, strongly coupled SJJ's.

The single fluxon in the stack of $N$ junctions is described by
$N$ characteristic lengths. With increasing velocity, magnetic
induction in a fluxon can change sign. For large $N$ there is no
Lorentz contraction of the fluxon at $u \rightarrow c_1$. This
results in absence of velocity matching behavior and a possibility
of "super-relativistic" fluxon motion with $u>c_1$. Such motion is
accompanied by generation of Josephson plasma waves moving along
with the fluxon ("Cherenkov" radiation) due to degeneration of
fluxon components with $c_m<u$.

Due to existence of multiple quasi-equilibrium fluxon
modes/submodes the state of the stack is not well defined for
given external conditions. It can only be described statistically
with a certain probability of being in any of quasi-equilibrium
states. This causes complicated and anomalous behavior of SJJ's:
(i) Characteristics of SJJ's exhibit strong fluctuations. (ii) The
$c$-axis critical current is multiple valued. The probability
distribution $P(I_c)$ has multiple maxima and $I_c(T,H)$ consist
of multiple sub-branches. (iii) The $I_c(H)$ patterns are
aperiodic. (iv) Flux-flow IVC's consist of multiple closely spaced
sub-branches.


\begin{thebibliography}{33}

\bibitem{LD} W.E.Lowrence and S.Doniach, in {\em Proc. LT-12, Kyoto, 1970}, ed.
by E.Kanada (Keigaku, Tokyo 1970), p.361

\bibitem{Muller}  R. Kleiner, F.Steinmeyer, G.Kunkel, P.M\"{u}ller, Phys.
Rev. Lett. {\bf 68}, 2394 (1992)

\bibitem{Mineev}  M.B.Mineev, G.S.Mkrtchyan, and V.V.Shmidt, J.Low Temp.
Phys. {\bf 45}, 497 (1981)

\bibitem{SBP}  S.Sakai, P.Bodin and N.F.Pedersen, J. Appl. Phys. {\bf 73},
2411 (1993)

\bibitem{charging_Tachiki} T.Koyama and M.Tachiki, Phys. Rev.B {\bf 54} (1996) 16183

\bibitem{charging_Preis} C.Preis, C.Helm, J.Keller, A.Sergeev and R.Kleiner,
Proc. SPIE {\bf 3480} (1998) 236

\bibitem{Ryndyk} D.A.Ryndyk, Phys.Rev.Lett. {\bf 80}, 3376 (1998)

\bibitem{Clem1}  J.Clem, Phys. Rev. B {\bf 43}, 8737 (1991)

\bibitem{Blatter}  G.Blatter, M.V.Feigelman, V.B.Geshkenbein, A.I.Larkin,
and V.M.Vinokur, Rev. Mod. Phys. {\bf 66}, 1125 (1994)

\bibitem{Nelson}  L.Balents and D.R.Nelson, Phys.Rev.B {\bf 52}, 12951 (1995)

\bibitem{IntPin}  M.Tachiki and S.Takahashi, Solid State Commun. {\bf 70},
291 (1989)

\bibitem{Bulaevskii73} L.N.Bulaevskii, Sov.Phys.JETP {\bf 37}, 1133 (1973)

\bibitem{Malomed} Yu.S.Kivshar and B.A.Malomed, Phys.Rev.B {\bf 37}, 9325
(1988)

\bibitem{Clem2}  J.Clem and M.Coffey, Phys. Rev. B {\bf 42}, 6209 (1990)

\bibitem{ClemCoffey91} J. R. Clem, M. W. Coffey, and Z. Hao, {\em Phys.Rev.B} {\bf 44},
2732 (1991).

\bibitem{Bul1} L.Bulaevskii and J.R.Clem, Phys.Rev.B {\bf 44}, 10234 (1991)

\bibitem{KrGol} V.M.Krasnov, N.F.Pedersen and A.A.Golubov, Physica C {\bf %
209}, 579 (1993)

\bibitem{Koshel} A.E.Koshelev, Phys. Rev. B {\bf 48}, 1180 (1993)

\bibitem{Klein1} R.Kleiner, Phys. Rev. B {\bf 50}, 6919 (1994)

\bibitem{Modes} V.M. Krasnov and D. Winkler, Phys. Rev. B {\bf 56}, 9106
(1997)

\bibitem{Shaf} S.E.Shafranjuk, M.Tachiki, and T.Yamashita Phys.Rev.B. {\bf %
57}, 13765 (1998)

\bibitem{Hu} X.Hu and M.Tachiki, Phys.Rev.Lett. {\bf 80}, 4044 (1998)

\bibitem{SubModes} V.M.Krasnov, N.Mros, A.Yurgens and D.Winkler, Physica C
{\bf 304}, 172 (1998)

\bibitem{SakPed} N.F.Pedersen and S.Sakai, Phys.Rev.B {\bf 58}, 2820 (1998)

\bibitem{SolPert} V.M. Krasnov, Phys. Rev. B {\bf 60}, 9313 (1999)

\bibitem{FluxonD} V.M. Krasnov and D.Winkler, Phys. Rev. B {\bf 60}, 13179 (1999)

\bibitem{Machida} M.Machida, T.Koyama, A.Tanaka, and M.Tachiki, Physica C {\bf 330}, 85 (2000)

\bibitem{PhysB} V.M.Krasnov, N.Mros, A.Yurgens, D.Winkler, Physica B {\bf 284-288}, 1856 (2000)

\bibitem{Vortex99} V.M.Krasnov, Physica C {\bf 332}, 308 (2000)

\bibitem{Compar} V.M.Krasnov, V.A.Oboznov, V.V.Ryazanov, N.Mros, A.Yurgens
and D.Winkler, Phys.Rev.B {\bf 61}, 766 (2000)

\bibitem{Kleiner62} R.Kleiner, T.Gaber, and G.Hechtfisher, Phys.Rev.B {\bf 62}, 4086 (2000)

\bibitem{FluxoN}  V.M. Krasnov, Phys. Rev. B {\bf 63}, 064519 (2001);

\bibitem{KrLaRy}  V.M.Krasnov, V.A.Larkin and V.V.Ryazanov, Physica C {\bf %
174}, 440 (1991)

\bibitem{Nakamura}  N.Nakamura, G.D.Gu, and N.Koshizuka, Phys.Rev.Lett {\bf %
71}, 915 (1993)

\bibitem{Zavar} M.Nider\"ost, R.Frassanito, M.Saalfrank,A.C.Mota, G.Blatter,
V.N.Zavaritsky, T.W.Li, and P.Kes, Phys.Rev.Lett {\bf %
81}, 3231 (1998)

\bibitem{YurgAPL}A.Yurgens, D.Winkler, T.Claeson, N.Zavaritsky,
Appl.Phys.Lett. {\bf 70} 1760 (1997)

\bibitem{Lee} J.U.Lee and J.E.Nordman, Physica C {\bf 277}, 7 (1997)

\bibitem{Latyshev} Yu.I.Latyshev, P.Monceau and V.N.Pavlenko, Physica C {\bf 293}, 174 (1997)

\bibitem{HechtfPRB} G.Hechtfisher, R.Kleiner, K.Schlenga, W.Walkenhorst, P.M\"uller and H.L.Johnson,
Phys.Rev.B {\bf 55}, 14638 (1997)

\bibitem{Irie} A.Irie, Y.Hirai, and G.Oya, Appl.Phys.Lett. {\bf 72}, 2159 (1998)

\bibitem{Fiske}  V.M.Krasnov, N.Mros, A.Yurgens, and D.Winkler, Phys.Rev.B {\bf 59}, 8463 (1999)

\bibitem{KrasT} V.M.Krasnov, A.Yurgens, D.Winkler, P.Delsing, and T.Claeson,
Phys. Rev. Lett. {\bf 84} (2000) 5860

\bibitem{KrasH} V.M.Krasnov,A.Kovalev,A.Yurgens,D.Winkler,
Phys.Rev.Lett. {\bf 86} (2001) 2657

\bibitem{Blamire} M.G.Blamire,E.Kirk,J.E.Evetts,T.M.Klapwijk, Phys.Rev.Lett. 66 (1991) 220.

\bibitem{Amin} H.Amin, M.G.Blamire and J.E.Evetts, IEEE Trans.Appl.Sup. 3, (1993) 2204

\bibitem{Kohlstedt} H.Kohlstedt, G.Hallmanns, I.P.Nevirkovets, D.Guggi, and C.Heiden, IEEE Trans. Appl. Supercond. 3, (1993) 2197.

\bibitem{Ust93} A.V.Ustinov, M.Cirillo, H.Kohlstedt, G.Hallmanns, C.Heiden and N.F.Pedersen, Phys.Rev.B 48 (1993) 10614

\bibitem{NbCuJ} V.M.Krasnov, N.F.Pedersen, V.A.Oboznov, and V.V.Ryazanov,
Phys.Rev.B {\bf 49}, 12969 (1994)

\bibitem{Nevirk} I.P.Nevirkovets, J.E.Evetts and M.G.Blamire, Phys.Lett.A C {\bf 187}, 119 (1994)

\bibitem{Monaco} R.Monaco, A.Polcari and L.Capogna, J.Appl.Phys. {\bf 78}, 3278 (1995)

\bibitem{Song} S.N.Song, P.R.Auvil, M.Ulmer and J.B.Ketterson, Phys.Rev.B {\bf 53}, R6018 (1996)

\bibitem{Costab} G.Carapella, G.Costabile, A.Petraglia, N.F.Pedersen, and J.Mygind,
Appl.Phys.Lett. {\bf 69}, 1300 (1996)

\bibitem{NbCuH}V.M.Krasnov,A.Kovalev,V.Oboznov,N.F.Pedersen
Phys.Rev.B {\bf 54}, 15448 (1996)

\bibitem{Thyssen} N.Thyssen, H.Kohlstedt, A.V.Ustinov,
IEEE Trans. on Appl. Sup. {\bf 7}, 2901 (1997); S.Sakai,
A.V.Ustinov, N.Thyssen, H.Kohlstedt, Phys. Rev. B {\bf 58},
5777 (1998)

\bibitem{HechtfPRL}  G.Hechtfisher, R.Kleiner, A.V.Ustinov, P.M\"{u}ller, Phys.
Rev. Lett. {\bf 79}, 1365 (1997)

\bibitem{Kirtley}
K.A.Moler, J.R.Kirtley, D.G.Hinks, T.W.Li, and M.Xu, Science {\bf
279},1193 (1998); A.A.Tsvetkov, et.al, Nature {\bf 395}, 360
(1998); J.R.Kirtley, V.G.Kogan, J.R.Clem, and K.A.Moler, Phys.
Rev. B {\bf 59}, 4343 (1999)

\bibitem{Laub} A.Laub, T.Doderer, S.G. Lachenmann, R.P.Huebener, and V.A.Oboznov
Phys.Rev.Lett. {\bf 75}, 1372 (1995)

\bibitem{Scott}  D.W.McLaughlin, and A.C.Scott, Phys.Rev.A {\bf 18}, 1652
(1978)

\bibitem{Goldob}  E.Goldobin, A.Wallraff and A.V.Ustinov, J.Low.Temp.Phys. {\bf 119}, (2000) 589

\bibitem{Levit}  L.S.Levitov, Phys. Rev. Lett. {\bf 66}, 224 (1991);

\bibitem{Watson}  G.I.Watson and G.S.Canright, Phys.Rev.B {\bf 48}, 15950
(1993)

\bibitem{Mros}  N.Mros, V.M.Krasnov, A.Yurgens, D.Winkler and T.Claeson,
Phys. Rev. B, {\bf 57}, R8135 (1998)

\bibitem{Schuller}  I.Banerjee and I.K.Schuller, J.Low.Temp.Phys. {\bf 54},
501 (1984)

\bibitem{Hc1ll}  V.M.Krasnov, A.E.Kovalev, V.A.Oboznov, and V.V.Ryazanov,
Physica C {\bf 215}, 265 (1993)

\bibitem{NbCuIc}  V.M.Krasnov, N.F.Pedersen, and V.A.Oboznov, Phys. Rev. B
{\bf 50}, 1106 (1994)

\bibitem{Logv}  P.Seng, R.Tidecks, K.Samwer, G.Yu.Logvenov, and V.A.Oboznov,
J.Low.Temp.Phys. {\bf 106}, 29 (1997)

\bibitem{Takahashi}  S.Takahashi and M.Tachiki, Phys.Rev.B {\bf 33}, 4620
(1986)

\bibitem{Barone}  A.Barone and G.Paterno, Physics and Applications of the
Josephson Effect (Wiley, New York, 1982)

\bibitem{Bul92}  L.N.Bulaevskii, J.R.Clem, and L.I.Glazman, Phys.Rev.B {\bf 46}%
, 350 (1992)

\bibitem{Kleiner94}  R.Kleiner, and P.M\"{u}ller, Phys. Rev. B {\bf 49}, 1327
(1994)

\bibitem{LatyshevPRL}  Yu.I.Latyshev, J.E.Nevelskaya, and P.Monceau,
Phys.Rev.Lett {\bf 77}, 932 (1996)

\bibitem{KleinSak} R.Kleiner, P.M\"{u}ller, H.Kohlstedt, N.F.Pedersen and S.Sakai,
Phys. Rev. B {\bf 50} (1994) 3942

\bibitem{Ling}  D.C.Ling, G.Yong, J.T.Chen, and L.E.Wenger, Phys.Rev.Lett. {\bf %
75}, 2011 (1995)

\bibitem{Lee2} J.U.Lee, P.Guptasarma, D.Hornbaker, A.ElKortas, D.Hinks, K.E.Gray,
Appl.Phys.Lett. {\bf 71}, 1412 (1997)

\bibitem{KrIEEE}  V.M.Krasnov, N.Mros, A.Yurgens, and D.Winkler, IEEE Trans.
on Appl. Supercond. {\bf 9}, 4499 (1999)

\bibitem{Shitov} S.V.Shitov, A.V.Ustinov, N.Iosad, and H.Kohlstedt, J.Appl.Phys. {\bf 80} (1996) 7134

\bibitem{Doh} Y.J.Doh, J.Kim, H.S.Chang, S.Chang, H.J.Lee, K.T.Kim, W.Lee, and J.H.Choy, Phys.Rev.B
{\bf 63} (2001) 144523





\end{thebibliography}
\end{document}